\documentclass[aps,prl,twocolumn,superscriptaddress]{revtex4-1}

\usepackage{amssymb}
\usepackage{natbib}

\usepackage{multirow}
\usepackage{bm}
\usepackage{amsmath}
\usepackage{graphicx}
\usepackage{epstopdf}
\usepackage{subfigure}
\usepackage{natbib}
\usepackage{epsfig}
\usepackage{amsfonts}
\usepackage{mathrsfs}
\usepackage{braket}
\usepackage{tabularx}
\usepackage[toc,page,title,titletoc,header]{appendix}
\usepackage{hyperref}
\hypersetup{colorlinks,allcolors=blue}

\usepackage{dsfont,amsthm,amsbsy}

\usepackage{fancyhdr}
\usepackage{lineno}
\begin{document}

\title{Tunable spin and valley excitations of correlated insulators in $\Gamma$-valley moir\'e bands}

\author{Benjamin A. Foutty}
\thanks{These authors contributed equally}
\affiliation{Geballe Laboratory for Advanced Materials, Stanford, CA 94305, USA}
\affiliation{Department of Physics, Stanford University, Stanford, CA 94305, USA}

\author{Jiachen Yu}
\thanks{These authors contributed equally}
\affiliation{Geballe Laboratory for Advanced Materials, Stanford, CA 94305, USA}
\affiliation{Department of Applied Physics, Stanford University, Stanford, CA 94305, USA}

\author{Trithep Devakul}
\thanks{These authors contributed equally}
\affiliation{Department of Physics, Massachusetts Institute of Technology, Cambridge, Massachusetts 02139, USA}

\author{Carlos R. Kometter}
\affiliation{Geballe Laboratory for Advanced Materials, Stanford, CA 94305, USA}
\affiliation{Department of Physics, Stanford University, Stanford, CA 94305, USA}

\author{Yang Zhang}
\affiliation{Department of Physics, Massachusetts Institute of Technology, Cambridge, Massachusetts 02139, USA}
\affiliation{Department of Physics and Astronomy, University of Tennessee, Knoxville, Tennessee 37996, USA}

\author{Kenji Watanabe}
\affiliation{Research Center for Functional Materials, National Institute for Material Science, 1-1 Namiki, Tsukuba 305-0044, Japan}

\author{Takashi Taniguchi}
\affiliation{International Center for Materials Nanoarchitectonics, National Institute for Material Science, 1-1 Namiki, Tsukuba 305-0044, Japan}

\author{Liang Fu}
\affiliation{Department of Physics, Massachusetts Institute of Technology, Cambridge, Massachusetts 02139, USA}

\author{Benjamin E. Feldman}
\email{bef@stanford.edu}

\affiliation{Geballe Laboratory for Advanced Materials, Stanford, CA 94305, USA}
\affiliation{Department of Physics, Stanford University, Stanford, CA 94305, USA}
\affiliation{Stanford Institute for Materials and Energy Sciences, SLAC National Accelerator Laboratory, Menlo Park, CA 94025, USA}


\begin{abstract}

    Moir\'e superlattices formed from transition metal dichalcogenides (TMDs) support a variety of quantum electronic phases that are highly tunable using applied electromagnetic fields. While the valley degree of freedom affects optoelectronic properties in the constituent TMDs, it has yet to be fully explored in moir\'e systems. Here, we establish twisted double bilayer WSe$_2$ as an experimental platform to study electronic correlations within $\Gamma$-valley moir\'e bands. Through local and global electronic compressibility measurements, we identify charge-ordered phases at multiple integer and fractional moir\'e fillings. By measuring the magnetic field dependence of their energy gaps and the chemical potential upon doping, we reveal spin-polarized ground states with spin polaron quasiparticle excitations. In addition, an applied displacement field induces a metal-insulator transition driven by tuning between $\Gamma$- and $K$-valley moir\'e bands. Our results demonstrate control over the spin and valley character of the correlated ground and excited states in this system.
    
\end{abstract}

\maketitle
\section{Introduction}

Control over the electronic band structure and internal quantum degrees of freedom in solid state materials is highly desirable to engineer novel functionality. Moir\'e van der Waals heterostructures, in which an interlayer twist and/or mismatch between lattice constants produces a long-wavelength spatial modulation, have attracted great interest due to their ability to host flat bands in which electronic interactions dominate over kinetic energy \cite{andrei_graphene_2020,balents_superconductivity_2020,kennes_moire_2021}. Moir\'e superlattices engineered from semiconducting transition metal dichalcogenides (TMDs) are a particularly flexible platform for investigating correlated electronic states because the resulting physics can be tuned by adjusting TMD composition and stacking orientation, while flat bands emerge without stringent constraints on twist angle \cite{wu_hubbard_2018,zhang_moire_2020,wu_topological_2019,angeli__2021,zhang_electronic_2021}. Multiple homo- and heterobilayer systems have been explored to date, revealing correlation-driven charge transfer and/or Mott insulating states at integer moir\'e fillings \cite{regan_mott_2020,tang_simulation_2020,wang_correlated_2020,li_charge-order-enhanced_2021,li_quantum_2021,xu_tunable_2022} and a series of generalized Wigner crystals at fractional fillings \cite{regan_mott_2020,xu_correlated_2020,chu_nanoscale_2020,li_charge-order-enhanced_2021,li_imaging_2021,huang_correlated_2021,tang_dielectric_2022}. These correlated electronic states are often strongly tuned by applied gate voltages and magnetic fields, which can favor different many-body ground states and/or modify the underlying single-particle bands \cite{ghiotto_quantum_2021,li_continuous_2021,li_quantum_2021,tang_dielectric_2022,zhang_pomeranchuk_2022}. As a result, the TMD platform is an ideal venue for quantum simulation and study of the Hubbard model, which is thought to capture the essential physics of conventional strongly correlated materials \cite{kennes_moire_2021,wu_hubbard_2018,tang_simulation_2020,pan_quantum_2020}. This has motivated careful measurement of the spin ordering of the ground states \cite{hu_competing_2021,xu_tunable_2022,wang_light-induced_2022,tang_evidence_2023}. However, experimental studies have primarily focused on low magnetic fields, and little is known about the nature of the lowest energy excitations or their field dependence.  

The valley degree of freedom, which captures the momentum of the low-energy states, also plays an important role in TMDs, and the valley character of the moir\'e bands derives from that of the constituent TMDs \cite{angeli__2021,magorrian_multifaceted_2021,vitale_flat_2021,xian_realization_2021}. If the moir\'e bands are localized around the $\Gamma$-valley, then these bands will be spin degenerate but have no valley degeneracy. This presents a qualitatively distinct physical system relative to $K$-valley TMDs, which have two degenerate spin-valley-locked bands \cite{angeli__2021,wang_correlated_2020}. Due to the structure of TMDs, there are also two distinct stackings of their moir\'e superlattices \cite{shabani_deep_2021}. Twisted $\Gamma$-valley moir\'e TMD systems with AA (i.e. near 3R) stacking are predicted to realize effective Hubbard models in honeycomb and Kagome  geometries \cite{angeli__2021,zhang_electronic_2021,xian_realization_2021}. AB (i.e. near 2H) stacked $\Gamma$-valley systems realize a triangular lattice, which can have flatter bands due to the larger effective mass at $\Gamma$, enhancing the importance of interactions. In untwisted TMDs, increasing the number of layers shifts the valence band edge from $K$ to $\Gamma$ \cite{mak_atomically_2010,xu_spin_2014}, and tuning the relative populations of these two valleys with applied electric field was demonstrated in magnetotransport measurements in trilayer WSe$_2$ \cite{movva_tunable_2018}. However, few experimental realizations of $\Gamma$-valley moir\'e systems have been reported, and their electronic properties are poorly characterized. The interplay of $\Gamma-K$-valley-tunable physics in the context of moir\'e bands also remains an open question in experiments. 

Here we study magnetic and electric displacement field dependence of correlated insulators in $\Gamma$-valley hole moir\'e bands in AB-stacked twisted double bilayer WSe$_2$ (tdWSe$_2$; Fig. 1a). We measure the chemical potential $\mu(n)$ and inverse electronic compressibility $\textrm{d}\mu/\textrm{d}n$ both locally using a scanning single-electron transistor (SET) and globally in dual gated devices. Our measurements reveal charge-ordered states at integer and fractional moir\'e filling factors $\nu = -1,-3$ and $-\frac{1}{3}$, where $\nu = -1$ corresponds to one hole per moir\'e unit cell. The measured gaps grow linearly with perpendicular magnetic field in the low-field regime, indicating spin-polarized ground states. The $\nu = -\frac{1}{3}$ gap saturates at high fields, and the behavior of the chemical potential in its vicinity provides evidence that itinerant spin polarons, bound states of added electrons and spin-flipped holes, are the lowest energy charged excitations \cite{davydova_itinerant_2022}. The thermodynamic energy gaps at integer fillings decrease monotonically with electric displacement field, driven by a single-particle band crossing between the lowest energy $\Gamma$ and $K$ moir\'e bands. Our experiments establish a new approach to engineer $\Gamma$-valley moir\'e bands and fully characterize the real- and momentum-space localization of the correlated ground states and their excitations. 

\begin{figure*}[t!]
    \renewcommand{\thefigure}{\arabic{figure}}
    \centering
    \includegraphics[scale =1.0]{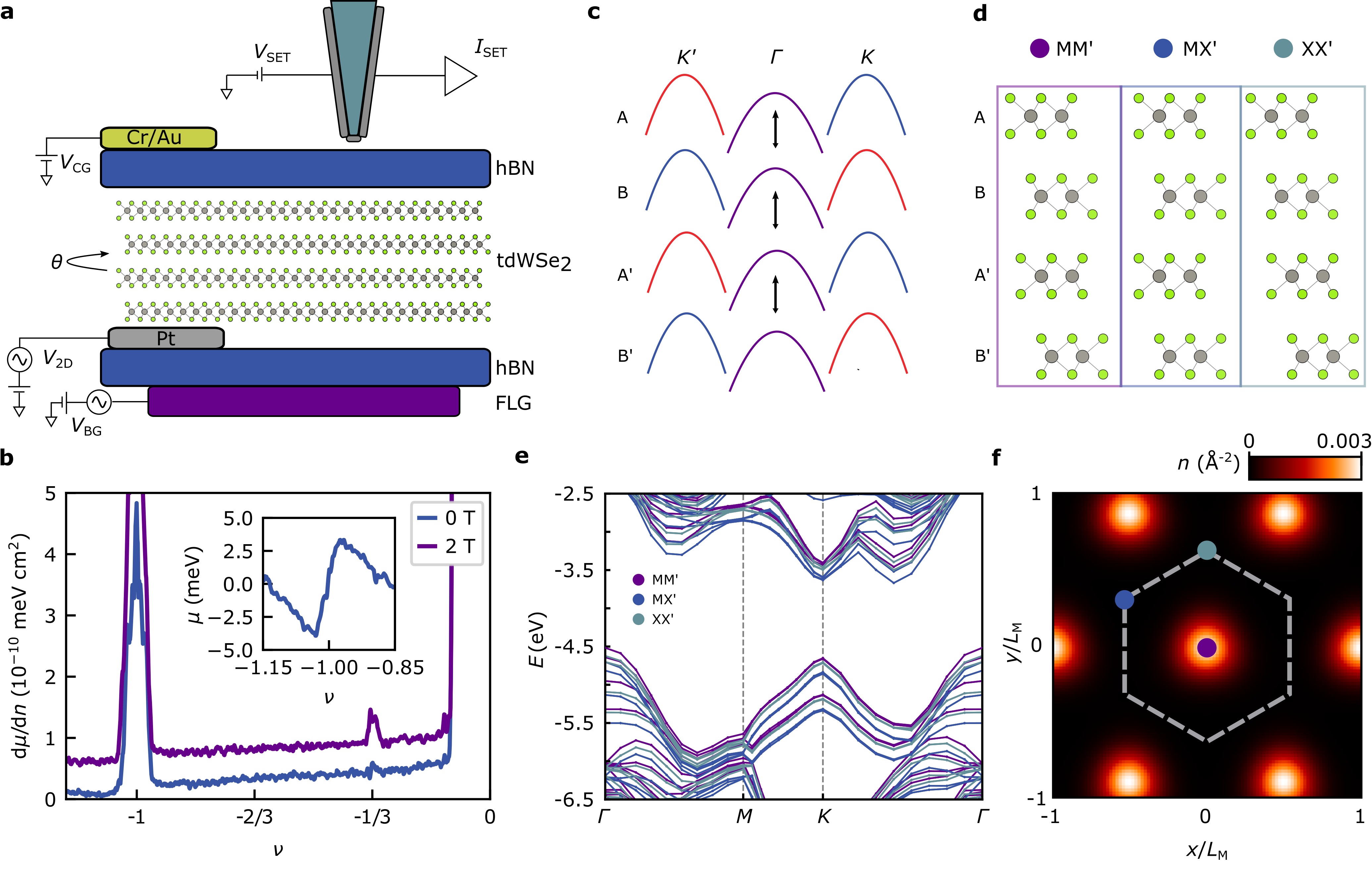}
    \caption{\textbf{$\Gamma$-valley moir\'e bands in twisted double bilayer WSe$_2$ (tdWSe$_2$)}. \textbf{a}, Schematic of device and measurement geometry. \textbf{b}, Local measurement of inverse electronic compressibility $\textrm{d}\mu/\textrm{d}n$ of Sample S1 as a function of moir\'e filling factor $\nu$ at magnetic fields $B=0$ T and $B = 2$ T (offset by $0.5\times 10^{-10}$ meV cm$^{2}$). The incompressible peak in $\textrm{d}\mu/\textrm{d}n$ is artificially enhanced in the a.c. measurement. Inset, simultaneous d.c. measurement of chemical potential $\mu(n)$ through the $\nu=-1$ gap. \textbf{c}, Schematic of valence band edges on each layer of AB-stacked tdWSe$_2$ before consideration of moir\'e bands. Red (blue) denotes spin up (down) at the $K$ and $K'$ points, while purple bands at $\Gamma$ are spin degenerate. Arrows show interlayer hybridization, which can occur at $\Gamma$ but is limited by spin mismatch at $K$ and $K'$. \textbf{d}, High symmetry stackings of tdWSe$_2$: $\textrm{MM}^\prime$, in which the W atoms are aligned vertically on the two internal layers of the 4-layer heterostructure, $\textrm{MX}^\prime$, in which all four layers realize $2H$ stacking, and $\textrm{XX}^\prime$, in which the Se atoms on the two internal layers are aligned vertically. \textbf{e}, Density functional theory calculated bands at the high symmetry points shown in \textbf{d}, with reference energy $E=0$ chosen to be the absolute vacuum level. The valence band edge occurs at the $\textrm{MM}^\prime$ site at $\Gamma$. \textbf{f}, Wannier orbitals of the lowest energy moir\'e band for 3.4$^\circ$ tdWSe$_2$ as calculated from a continuum model on the bands shown in \textbf{e}. States are well-localized at the $\textrm{MM}^\prime$ sites (purple dot). There is no charge density on the other high-symmetry sites $\textrm{MX}^\prime$ (blue) or $\textrm{XX}^\prime$ (teal).} 
    \label{fig:Fig1}
\end{figure*}

\section{$\Gamma$-valley moir\'e bands in \lowercase{td}WS\lowercase{e}$_2$}

We focus first on local measurements using a scanning SET, schematically illustrated in Fig. 1a. The SET simultaneously and independently probes two related thermodynamic quantities, the inverse electronic compressibility d$\mu$/d$n$ at low a.c. frequency and the chemical potential $\mu(n)$ on d.c. timescales (Methods). This technique mitigates the problem of large contact/intrinsic resistance that often complicates electrical measurements in TMD moir\'e materials \cite{regan_mott_2020,li_charge-order-enhanced_2021}. In Sample S1 (Fig. 1b), we observe a pronounced incompressible state at $n \approx 3.9\times$ 10$^{12}$ cm$^{-2}$, which we identify with $\nu = -1$. A weaker incompressible peak is also evident at $\nu = -\frac{1}{3}$, and it becomes prominent in a small applied out of plane magnetic field $B$. Both of these features are generically present across a wide range of spatial locations (Supplementary Sec. 1), and from the densities at which they occur, we extract a twist angle of $3.4^\circ$ (Methods). The corresponding thermodynamic gap $\Delta_\nu$, defined as the step size in $\mu(n)$ at filling $\nu$, can be determined from the d.c. measurement of chemical potential $\mu(n)$ (Fig. 1b, inset), yielding $\Delta_{-1} \approx 6$ meV at $B=0$ (Supplementary Sec. 2). We also measured a similar gap at $\nu = -1$ in a  second device (Sample S2) with a smaller twist angle of $2.6^\circ$. Both samples show very little ($<0.1^\circ$) twist angle variation (Supplementary Sec. 1), so we focus on a single spatial location in each device throughout the main text. 

As a first step toward understanding the correlated states, we consider the low energy valence band structure at the single-particle level in AB-stacked tdWSe$_2$. Due to strong spin-orbit coupling in WSe$_2$, interlayer coupling between adjacent layers is suppressed by spin mismatch at the the $K$ and $K'$ valleys \cite{zhu_giant_2011,xu_spin_2014,shi_bilayer_2022}. At the $\Gamma$ point, there is no such barrier to interlayer tunneling and hybridization can increase the valence band energy at $\Gamma$ as the layer number increases (Fig. 1c). This has been both predicted in theoretical calculations and observed in experiments in few-layer WSe$_2$ \cite{liu_electronic_2015,wilson_determination_nodate, movva_tunable_2018}. Additionally, a previous ARPES measurement of tdWSe$_2$ found the valence band maximum to occur at $\Gamma$ rather than $K$ \cite{an_interaction_2020}.

We confirm this picture by performing density functional theory (DFT) calculations at zero displacement field on the high symmetry stackings of tdWSe$_2$ (Fig. 1d, Methods). The respective band structures (Fig. 1e) demonstrate that the valence band maximum occurs at the $\Gamma$ valley in the $\textrm{MM}^\prime$ stacking and is separated by roughly 100 meV from the energies at other stackings. Using the energies obtained from DFT, we develop an effective continuum model describing moir\'e bands from both $\Gamma$ and $K$ valley bands involving all four layers (Supplementary Sec. 3). This continuum model also allows us to later incorporate the effect of a perpendicular displacement field as an effective layer potential difference. The combination of large effective mass at $\Gamma$ and strong moir\'e potential results in a very flat first moir\'e band which is well localized at the $\textrm{MM}^\prime$ sites (Fig. 1f, Supplementary Sec. 3). Due to the two-fold spin degeneracy of the moir\'e bands, the experimentally observed gap at $\nu = -1$ must be driven by interactions. Measurements as a function of displacement field, discussed in detail below, suggest that the lowest-energy charge excitations at $\nu = -1$ populate a moir\'e band at the $K$-valley. Combining this experimental finding with first-principles calculations that take into account Coulomb and charge transfer energies, we find that this $K$-valley moir\'e band will be well-localized at $\textrm{XX}^\prime$ in real space (Supplementary Sec. 3).

\begin{figure*}[t]
    \renewcommand{\thefigure}{\arabic{figure}}
    \centering
    \includegraphics[scale =1.0]{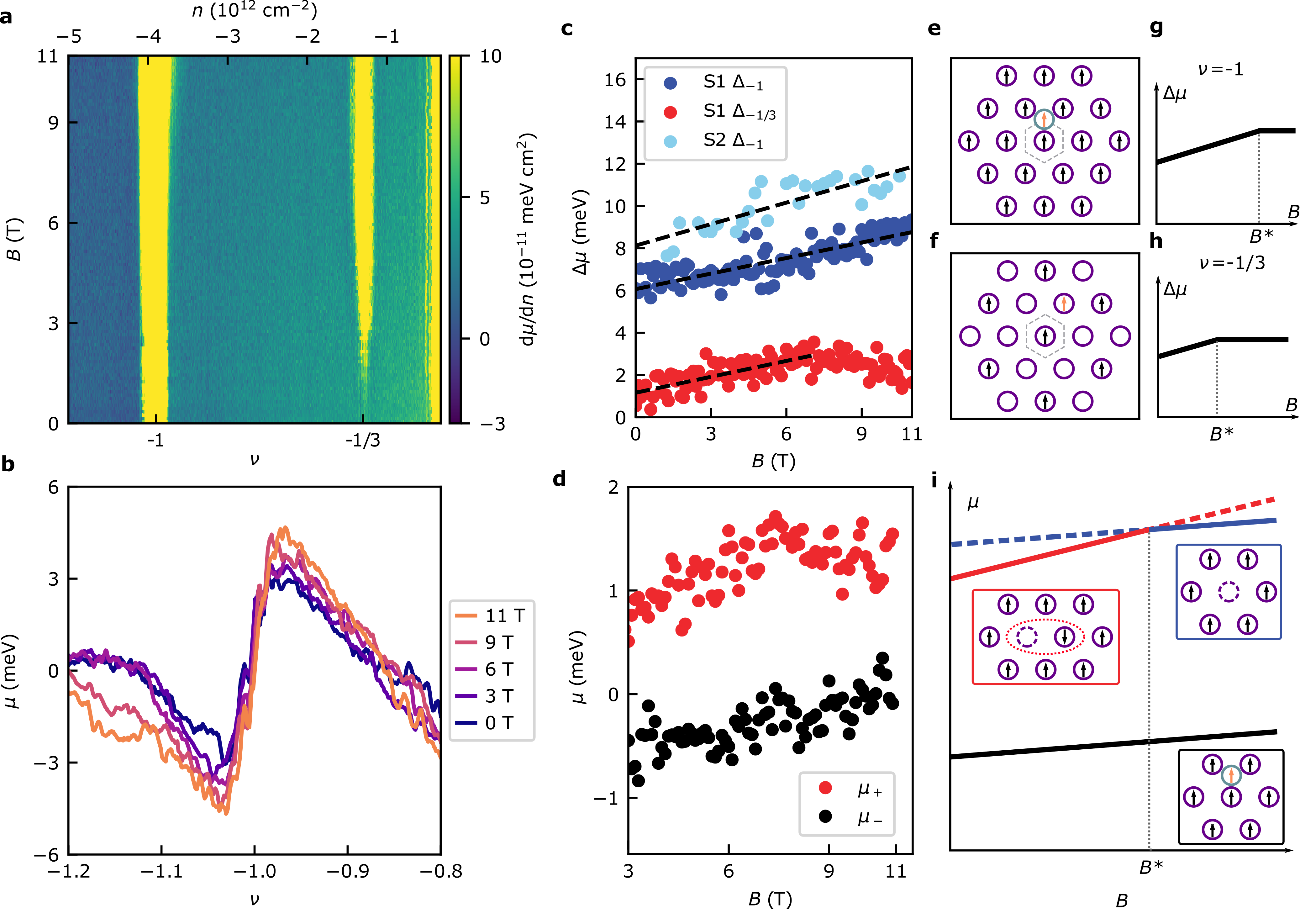}
    \caption{\textbf{Magnetic field dependence of correlated insulating ground states and their excitations}. \textbf{a}, d$\mu$/d$n$ as a function $\nu$ (bottom axis), density $n$ (top axis), and of out-of-plane magnetic field $B$. \textbf{b}, Linecuts of d.c. measured $\mu(\nu)$ around $\nu = -1$ at various magnetic fields. \textbf{c}, Extracted charge gaps at $\nu = -1$, $\nu = -\frac{1}{3}$ in Sample S1 and $\nu = -1$ in Sample S2. Dashed lines show linear fits to the data. All increase at low field, while $\Delta_{-1/3}$ saturates and decreases slightly beginning at $B = 7$ T. \textbf{d}, Chemical potential $\mu_+$ and $\mu_-$ measured at $\nu = -\frac{1}{3} \pm \epsilon$ for $\epsilon > 0$ on either side of the measured gap. On the lower hole doping side of the gap, ($\mu_+$), the slope is discontinuous in $B$, whereas at higher hole doping the behavior is smooth. \textbf{e-f}, Schematic of real space spin ordering at  $\nu =-1$ (\textbf{e}) and $\nu=-\frac{1}{3}$ (\textbf{f}). Ground state hole spins are shown in black, while the lowest energy hole excitations are shown in orange. \textbf{g,h}, Schematic of the expected gaps as a function of $B$. At low fields, the gap should increase due to the creation of spin polarons on the lower hole doping side of the gap. Above a critical field $B^*$,, spin polarons are no longer favored and the gap is constant in $B$. This crossover point should be higher at $\nu = -1$ due to higher effective hopping $t$. \textbf{i}, Schematic of chemical potential of states at half filling of a charge transfer triangular lattice as a function of magnetic field $B$. Black line and inset denote the lowest energy excitation at higher hole density, adding an aligned hole at an interstitial. Red line and inset denote the spin polaron excitation at lower hole density, which is favored at intermediate fields $B < B^*$, while the blue line and inset denote the excitation of removing a single hole. Solid circles with spins denote holes, while dashed circles denote the absence of a hole.}
    \label{fig:Fig2}
\end{figure*}

\section{Magnetic field dependence of charge-ordered states}

The evolution of the charge gaps in a perpendicular magnetic field clarifies the spin physics of the correlated insulating ground states and their charged excitations. Both the $\nu=-\frac{1}{3}$ and $\nu=-1$ gaps grow in a magnetic field (Fig. 2a-b, Supplementary Sec. 2). $\Delta_{-1}$ grows linearly with $B$ throughout the experimentally accessible field range, whereas $\Delta_{-\frac{1}{3}}$ increases up to $B=7$ T before saturating or decreasing slightly at high fields (Fig. 2c). The regime of linear growth reflects Zeeman energy shifts, indicating that the insulating state is spin-polarized and a spin-flip is involved in the lowest energy particle-hole excitation. This is consistent with prior reports of small exchange coupling $J$ in moiré systems, such that a small magnetic field is sufficient to polarize the spins in the gapped ground state \cite{tang_simulation_2020}. However, we note that the observation of spin-polarized correlated insulators in a magnetic field differs from reports in twsited bilayer WSe$_2$ \cite{ghiotto_quantum_2021}. We assume that the insulating state at $\nu = -\frac{1}{3}$ is a $\sqrt{3} \times \sqrt{3}$ generalized Wigner crystal, so that both of these states realize triangular lattices in real space (Fig. 2e-f).

The saturating gap at $\nu=-\frac{1}{3}$ indicates a change in the nature of the excitations at high fields. The thermodynamic gap that we measure is equivalent to the particle-hole excitation energy (Supplementary Sec. 4), and therefore requires us to take into account the magnetic field dependence of the excitations both above and below the gap. One possible explanation is that the lowest energy excitation just below the gap is the addition of a hole of opposite spin at low fields, before Zeeman coupling overcomes exchange to favor adding an aligned hole (Supplementary Sec. 5). In this scenario, the excitation above the gap would be the removal of a single hole, independent of magnetic field.

However, recent theory suggests that more exotic spin polaron quasiparticles may be relevant in a half-filled triangular lattice with a spin-polarized ground state \cite{davydova_itinerant_2022}. In this second scenario, the excitation below the gap remains the same at all magnetic fields: addition of an aligned hole at an interstitial (charge-transfer) site (Fig. 2i, black inset). Above the gap, the lowest energy excitation below a critical field $B^*$ involves removing a hole and flipping the spin of an adjacent remaining hole on the lattice, creating an itinerant spin polaron quasiparticle (Fig. 2i, red inset) with effective spin $\frac{3}{2}$. At magnetic fields exceeding $B^*$, formation of spin polarons is no longer favored due to Zeeman coupling, and the excitation involves removing a bare hole without flipping any other spins (Fig. 2i, blue inset).

To distinguish between these two pictures, both of which predict a saturation of the gap size, we examine the behavior of the chemical potential as we dope away from the charge ordered state. The first scenario would involve a sharp change in the behavior of the chemical potential $\mu_-$ below the gap (Supplementary Sec. 5). In contrast, the spin polaron scenario would lead to a change in the chemical potential above the gap, $\mu_+$: the theoretical chemical potentials for each excitation are shown in Fig. 2i. In Fig. 2d, we show the magnetic field dependence of the experimentally measured chemical potentials $\mu_+$ and $\mu_-$ on either side of the $\nu = -\frac{1}{3}$ gap (Supplementary Sec. 6). While $\mu_-$ is completely smooth, $\mu_+$ changes slope sharply around $B = 7$ T, consistent only with the spin polaron picture. We therefore conclude that spin polarons are favored at lower hole doping of the $\nu = -\frac{1}{3}$ insulator below $B = 7$ T, providing the first experimental evidence for these composite quasiparticles.

Theoretically, the critical field $B^*$ scales with the hopping $t$ on the triangular lattice \cite{davydova_itinerant_2022}. Because the sites are closer together for the insulating state at $\nu=-1$ as compared to the generalized Wigner crystal at $\nu = -\frac{1}{3}$, the effective $t$ will be higher at $\nu=-1$, such that the crossover field $B^*$ at $\nu = -1$ is inaccessible in our measurements (Fig. 2e-h). Nonetheless, the observed chemical potential evolution on either side of the $\nu = -1$ gap suggests that the spin polaron excitations are relevant there as well (Supplementary Sec. 6).

Finally, we fit the linear growth of the $\nu = -1$ and $\nu = -\frac{1}{3}$ gaps in a magnetic field with an effective $g$-factor $g^*$, defined by $\Delta_\nu(B) = \Delta_\nu(0) + g^* \mu_B B$. In Sample S1, both gaps at $\nu=-1$ and $\nu=-\frac{1}{3}$ exhibit a similar effective $g$-factor $g^* = 4.3 \pm 0.8$. In Sample S2, we measure a larger $g^* = 5.8 \pm 1.5$ for $\nu=-1$. These values exceed the bare spin contribution $g_s = 2$ that we would expect from the predicted excitation (Supplementary Sec. 4). At $\nu = -1$, this likely reflects additional contributions from orbital effects \cite{aivazian_magnetic_2015,movva_tunable_2018, gustafsson_ambipolar_2018}. At $\nu = -\frac{1}{3}$, we do not expect an orbital component as the gap does not involve any change in moir\'e band. However, exchange enhancement of $g^*$ has been observed previously in TMDs and two-dimensional electron gases (2DEGs), particularly at low densities, and the observation of $g^* > 2$ at $\nu =-\frac{1}{3}$ likely stems from that effect \cite{gustafsson_ambipolar_2018,movva_density-dependent_2017,nicholas_exchange_1988,tutuc_spin_2002}.

Away from the gapped states, our measurements also reveal regions with negative d$\mu$/d$n$. This is visible on either side of the gap at $\nu = -1$ in Fig. 2b, and occurs over a wide range of fillings, a signature of strong Coulomb interactions \cite{eisenstein_negative_1992,eisenstein_compressibility_1994}. This negative compressibility can be understood through consideration of exchange interactions, which are sizable relative to the density of states contribution to $\mu(n)$ in this system  (see Supplementary Secs. 7-8 for further discussion). The overall size of negative compressibility increases in a magnetic field, suggesting paramagnetism through much of the density range outside of the correlated insulator gaps.

\begin{figure*}[t]
    \renewcommand{\thefigure}{\arabic{figure}}
    \centering
    \includegraphics[scale=1.0]{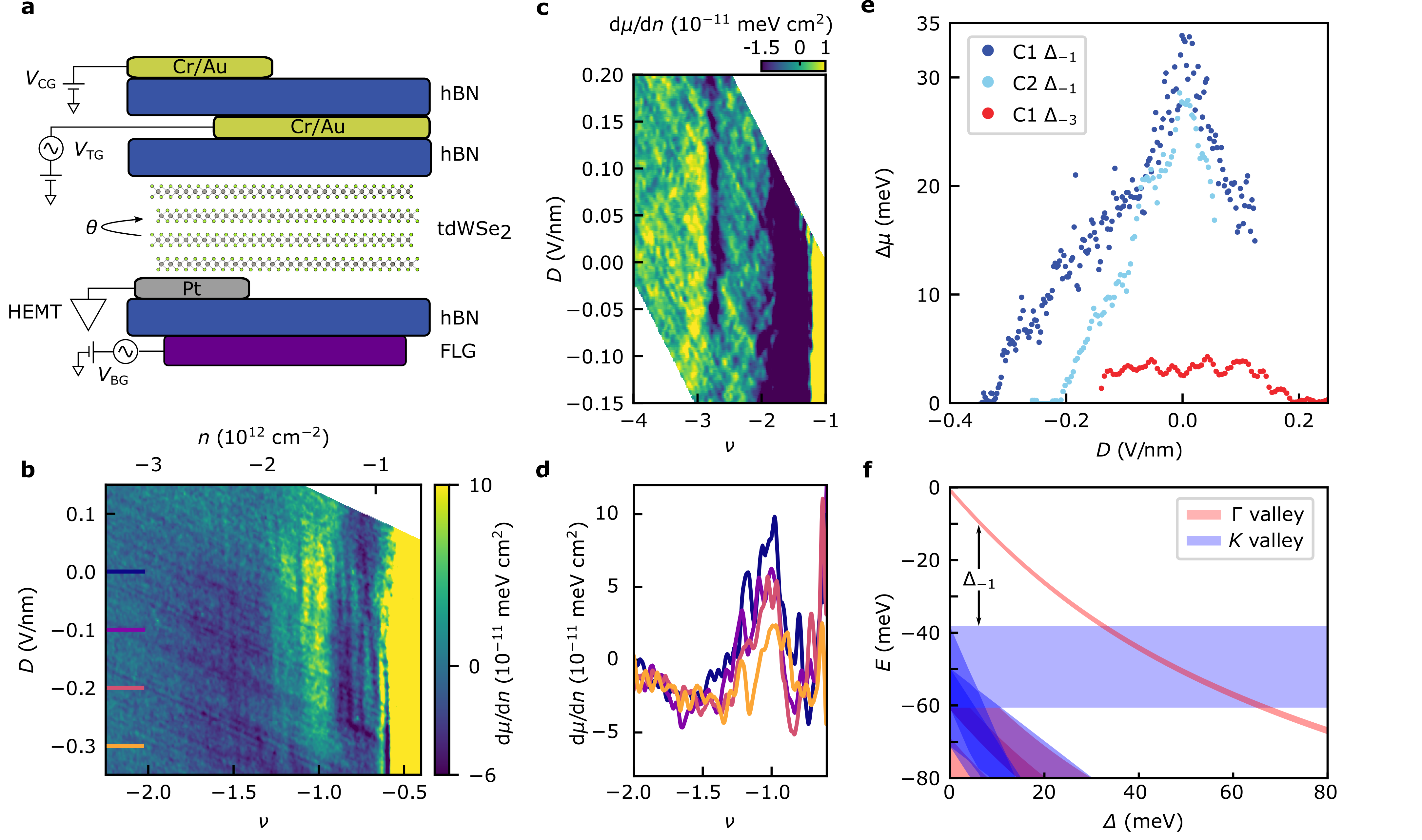}
    \caption{\textbf{Displacement field dependence of correlated insulators}. \textbf{a}, Schematic of a dual-gated capacitance device. \textbf{b} d$\mu$/d$n$ in Sample C1 as a function of $\nu$ and displacement field $D$ as extracted from top gate capacitance. \textbf{c}, d$\mu$/d$n$ around $\nu = -3$ in Sample C1. \textbf{d}, Linecuts of $\textrm{d}\mu/\textrm{d}n$ at various displacement fields from the data in panel \textbf{b}. \textbf{e}, Measured gaps $\Delta \mu$ at $\nu = -1$ from Samples C1 and C2, with twist angles $\theta = 2.2^\circ$ and $\theta = 3.1^\circ$, and gap at $\nu = -3$ in Sample C1. The gap size doesn't depend strongly on the twist angle. \textbf{f}, Continuum model predictions of dependence on moir\'e bands from the $\Gamma$- and $K$-valleys on layer potential difference $\Delta$, defined such that the energy of the topmost layer is constant as $\Delta$ is increased. Widths of the shaded regions show the bandwidth of respective moir\'e bands. At low layer potential difference, the lowest energy states are from the $\Gamma$ valley, whereas at large layer potential difference, the lowest energy hole state is localized on the topmost layer at the $K$ valley.} 
    \label{fig:Fig3}
\end{figure*}

\section{Displacement field tuning of moir\'e bands}
We next discuss the displacement field dependence of the correlated insulators, which has previously been shown to induce particularly strong effects in homobilayers where states are extended across layers \cite{wang_correlated_2020,xu_tunable_2022}. To study this thermodynamically in tdWSe$_2$, we fabricate samples with top and bottom gates and measure the sample-gate capacitance as a function of moir\'e filling factor and electrical displacement field $D$ (Fig. 3a, Methods). We studied two samples (C1 and C2) using this technique, with twist angles of $2.2^{\circ}$ and $3.1^{\circ}$, respectively. As hole density is increased beyond the mobility edge at $|n| \approx 0.8 \times 10^{12} \textrm{cm}^{-2}$, a broad region of negative $\textrm{d}\mu/\textrm{d}n$ is present before an insulating gap appears at $\nu = -1$ (Fig. 3b, Supplementary Sec. 9). In both samples, $\Delta_{-1}$ reaches a maximum around 30 meV at $D=0$ and closes monotonically with $|D|$, disappearing around $|D| \approx 0.2-0.4$ V/nm (Fig. 3d-e). Despite the large apparent difference in gap sizes, this is largely consistent with the SET measurements, where the single gate geometry leads to a nonzero displacement field (Supplementary Sec. 10). Qualitatively similar behavior occurs at $\nu = -3$ (Fig. 3c), whose gap also vanishes at high displacement field but shows a plateau within the signal to noise for $|D| < 0.1 $ V/nm (Fig. 3e) 

Previous works in TMD moir\'e systems have shown that displacement fields can drive a continuous metal-insulator transition by changing the bandwidth of the underlying moir\'e bands \cite{ghiotto_quantum_2021,li_continuous_2021}. In those experiments, the transport gap closes over a small range of displacement field ($D < 0.1$ V/nm) and does not vary strongly with $D$ outside of that range. At $\nu=-1$ in our device, the gap closes smoothly over a much larger range of $D$, and changes continuously all the way to $D = 0$. Simultaneously, our continuum modeling suggests the bandwidth of the $\Gamma$ moir\'e bands should be largely unaffected by displacement field.

Instead, our measurements and modeling are consistent with the lowest energy moir\'e band switching from $\Gamma$ to $K$ at sufficient layer potential imbalance (Fig. 3f, Supplementary Sec. 5). Because the $\nu = -1$ gap is set by the charge transfer excitation from $\Gamma$ to $K$ moir\'e bands, tuning the relative moir\'e band energies leads to a continual change in the measured gap. Past the critical displacement field where the $K$-valley moir\'e band crosses, we do not measure an incompressible state at $\nu=-1$, likely due to band overlap and/or the greater dispersion of the $K$-valley moir\'e band. The measured gap sizes and critical displacement field at $\nu=-1$ show weak dependence on twist angle: Samples C1 and C2 have twist angles that differ by about $1^\circ$ but the extracted gap at $D=0$ is similar across devices. This is because the gap size and displacement field dependence is largely set by the relative energy of the valence band edge of $\Gamma$ and $K$ at the band edge (Fig. 1e) and the effect that displacement field has on those energies, rather than details of the moir\'e band structure itself. The continuous closing of the gap from its maximum at $D=0$ provides experimental evidence that the lowest energy moir\'e bands at low displacement fields are from the $\Gamma$-valley and that even at $D=0$, the lowest energy excitations at $\nu = -1$ are to $K$-valley moir\'e bands. 

\section{Outlook}
In conclusion, we show prominent effects of electronic interactions in $\Gamma$-valley moir\'e bands over a wide range of parameter space in tdWSe$_2$. The correlated insulators we observe exhibit distinct magnetic and displacement field dependence compared to reports of other TMD moir\'e superlattices. Measurements of the chemical potential upon doping these charge-ordered states reveal evidence for spin polaron quasiparticle excitations. Together, this demonstrates that AB-stacked $\Gamma$-valley moir\'e TMDs realize novel parameter regimes of a mesoscopic triangular lattice model. Our work also suggests a clear path toward engineering low-disorder $\Gamma$-valley bands to look at new lattice geometries including honeycomb, Kagome, and anisotropic multi-orbital generalizations of the Hubbard model \cite{angeli__2021}, as well as spontaneous ferroelectric states \cite{zhang_electronic_2021}. A relatively modest displacement field is sufficient to tune the valley character of the lowest TMD moir\'e band from $\Gamma$ to $K$, providing a way to electrically switch between dramatically different moir\'e band structures within a single device. Further, our d.c. thermodynamic sensing modality enables quantitative measurement of TMD moir\'e systems down to low densities. This paves the way to study TMD moir\'e superlattices with longer wavelengths where the relevant experimental carrier density ranges become smaller, but electronic correlations are even stronger relative to bandwidth.



\section{Methods}
\subsection{Sample fabrication}
The tdWSe$_2$ devices were fabricated using standard dry transfer techniques. An exfoliated Bernal bilayer WSe$_2$ flake (sources: 2DSemiconductors, HQGraphene) was pre-cut by a conductive AFM probe in contact mode, with an a.c. excitation of 10 V at 50 kHz in order to facilitate the stacking process and alleviate strain. Using a poly(bisphenol A carbonate) (PC)/polydimethylsiloxane (PDMS) stamp, we pick up a thin (15-30 nm) hexagonal boron nitride (hBN) flake, followed by the first half of the bilayer WSe$_2$ flake, and then the second half rotated to a controlled angle of $2^\circ - 3.5^\circ$. Separately, we prepare a stack with a bottom hBN (25-40 nm thick) and a graphite (5-10 nm) back gate, on which we deposit pre-patterned Cr/Pt contacts (6-12 nm). This is annealed at $\approx 300\ ^\circ$C overnight to clean polymer and resist residues before depositing the tdWSe$_2$ stack on the Pt contacts. For samples S1 and S2, local Cr/Au ``contact'' gates (3nm/50 nm) were patterned above the Cr/Pt contacts after assembly in order to locally dope the contact regions so that they achieve Ohmic contact. For Samples C1 and C2, an additional Cr/Au top gate is first patterned over the device region (not the Pt contacts), after which a secondary hBN is set down on top of the device, followed by Cr/Au gates over the contacts. We used standard e-beam lithography techniques to fabricate contacts and top gates.

\subsection{Density and twist angle determination}
For Samples S1 and S2, both the band edge and hBN dielectric capacitance are determined by fitting Landau level oscillations (Supplementary Sec. 11). The band edge is taken to be the point to which these oscillations extrapolate at $B=0$, and the fitted slopes give the sample-gate capacitance, which specifies the conversion between gate voltage and hole density. In both samples, the location in gate voltage where we see effective charging of the TMD sample are a few $10^{11}\ \textrm{cm}^{-2}$ more doped than the band edge taken from the Landau fan. We attribute this difference to a mobility edge at low TMD densities \cite{wang_correlated_2020}. Based on the density of the $\nu = -1$ gapped state, we convert from the density of one hole per moir\'e unit cell $n_s$ to an angle $\theta$ by the relation $\frac{1}{n_s} = \frac{\sqrt{3}a}{4-4\cos(\theta)}$ where $a = .328$ nm is the lattice constant of WSe$_2$. Twist angles measured using this method agree to within ($\pm 0.3^\circ$) of the targeted rotation during the fabrication process, and are also confirmed to within $(\pm 0.5^\circ)$ based on edge orientation as measured from optical images and using atomic force microscopy. Further details and support for the twist angle assignment for each sample are provided in Supplementary Sec. 11.

For Samples C1-C2, we do not have magnetic field dependence to fix the band edge. The capacitances are estimated from hBN thickness and average measured dielectric constants for our batch of hBN samples, $\epsilon \approx 3.1$. The location of the band edge in gate voltage is determined by the spacing of the $\nu=-1$ and $\nu=-3$ features (in Sample C1) and lower frequency/higher contact gate capacitance linetraces where the sample charges better closer to the band edge (Supplementary Sec. 11). 

\subsection{SET measurement}
The SET sensor was fabricated by evaporating aluminum onto a pulled quartz rod, with an estimated diameter at the apex of $ 50 - 100$ nm. The SET ``tip" is brought to about $50$ nm above the sample surface. Scanning SET measurements were performed in a  Unisoku USM 1300 scanning probe microscope with a customized microscope head. a.c. excitations of order 5-10 mV were applied to both sample and back gate at distinct frequencies between 200 and 900 Hz. We then measure inverse compressibility $\textrm{d}\mu/\textrm{d}n \propto I_{\textrm{BG}} / I_{\textrm{2D}}$ where $ I_{\textrm{BG}}$ and $I_{\textrm{2D}}$ are measurements of the SET current demodulated at respective frequencies of the back gate and sample excitations \cite{yu_correlated_2022}. A d.c. offset voltage $V_{\textrm{2D}}$ is applied to the sample to maintain the working point of the SET at its maximum sensitivity point within a Coulomb blockade oscillation fringe chosen to be near the ``flat-band'' condition where the tip does not gate the sample. This minimizes tip-induced doping and provides a direct measurement of $\mu(n)$. The contact gates are held at a large, negative voltage throughout the measurement. All SET measurements are taken at $T = 330$ mK.

\subsection{Capacitance measurement in dual-gated devices}
We use standard high electron mobility transistor (HEMT) techniques to measure the device capacitance \cite{shi_odd-_2020,li_charge-order-enhanced_2021}. An FHX35X transistor is glued adjacent to the sample with the HEMT gate connected to the sample contacts. The sample voltage $V_{2D}$ is held fixed, setting the HEMT gain, while a small current is sourced between the source and drain leads of the transistor. Throughout the measurement, the contact gates are held fixed at a large, negative voltage $V_{\textrm{CG}}$ to maintain ohmic contact resistance independent of the applied gate voltages. The d.c. voltages $V_{\textrm{BG}}$ and $V_{\textrm{TG}}$ applied to top and back gate, respectively, independently tune the density $n$ and displacement field $D$. We define the displacement field as $D = \frac{1}{2\epsilon_0} (c_t(V_{\textrm{TG}} - \phi_0) - c_b(V_{\textrm{BG}} - \phi_1))$ where $c_{t(b)}$ is the capacitance of the top (bottom) hBN dielectric and $\phi_{0(1)}$ is the estimated work function difference between our TMD sample and gates. We take $\phi_1 = -0.8$ V and $\phi_0 = -0.1$ V, as checked by SET measurements in separate samples, which are sensitive to the work function. a.c. excitation voltages are applied to the top and back gates with amplitude 5-10 mV and frequency ranging from 300 Hz - 1 kHz, and the amplified capacitance (and out-of-phase dissipation) signal is measured by the a.c. voltage drop across the HEMT leads. All capacitance measurements presented in the main text were taken at $T = 4.2$ K (see Supplementary Sec. 12 for temperature dependence up to $T \approx 15$ K). To convert from measured signals to quantitative units, we normalize the signal using the known device capacitance and the difference in the measured signal between the fully gapped and highly doped regions of the sample, before applying a lumped circuit model to convert from the capacitive and dissipative components to d$\mu$/d$n$ (Supplementary Sec. 13). Dual gated devices are fabricated on an undoped Si/SiO$_2$ substrate to avoid parasitic capacitances, and the devices were etched (CHF$_3$/O$_2$ plasma etching) around the top gate to minimize singly gated regions of sample. 

\subsection{Density functional calculation}
Density functional calculations are performed using generalized gradient approximation with SCAN+rVV10 Van der Waals density functional \cite{peng_versatile_2016}, as implemented in the Vienna Ab initio Simulation Package \cite{kresse_efficient_1996}. Pseudopotentials are used to describe the electron-ion interactions. We construct the moir\'e band structure of tdWSe$_2$ by calculating the band structures of untwisted double bilayers with $\textrm{MM}^\prime$, $\textrm{MX}^\prime$ and $\textrm{XX}^\prime$ stackings and ``stitching'' the results together with the continuum model approach. The vacuum spacing is larger than 20 \AA{} to avoid artificial interaction between the periodic images along the z direction. The structure relaxation is performed with force on each atom less than 0.01 eV/A. We use $12\times 12 \times 1$ for structure relaxation and self-consistent calculation. The more accurate SCAN+rVV10 van der Waals density functional gives the relaxed layer distances as 6.64 \AA{}, 6.62 \AA{} and 7.10 \AA{} for $\textrm{MM}^\prime$, $\textrm{MX}^\prime$ and $\textrm{XX}^\prime$ stacking structures, respectively. By calculating the work function from electrostatic energy of converged charge density, we plot in Fig. 1e the band structure of $\textrm{MM}^\prime$, $\textrm{MX}^\prime$ and $\textrm{XX}^\prime$-stacked double bilayers, with reference energy $E=0$ chosen to be the absolute vacuum level. 

Our DFT calculation finds that the valence band maximum of tdWSe$_2$ is  at $\Gamma$ and lies roughly 100 meV above the band edge at $K$, consistent with a recent angle-resolved photoemission spectroscopy measurement \cite{an_interaction_2020}. Unlike monolayer and bilayer, the reversed ordering of $\Gamma$ and $K$ band edges in four-layer WSe$_2$ results from the strong interlayer tunneling around $\Gamma$, which leads to large energy splittings on the order of $0.1$ eV between layer-hybridized $\Gamma$-valley bands.

\section{Data availability}
The data that supports the findings of this study are available from the corresponding authors upon reasonable request.

\section{Code availability}
The codes that support the findings of this study are available from the corresponding authors upon reasonable request.

\section{Acknowledgements}
We acknowledge helpful conversations with Allan H. MacDonald. We thank Tony Heinz, Aidan O'Beirne, and Henrique Bucker Ribeiro for their assistance with SHG measurements. Experimental work was primarily supported by NSF-DMR-2103910. B.E.F. acknowledges an Alfred P. Sloan Foundation Fellowship and a Cottrell Scholar Award. The work at Massachusetts Institute of Technology is supported by a Simons Investigator Award from the Simons Foundation. L.F. is partly supported by the David and Lucile Packard Foundation. K.W. and T.T. acknowledge support from JSPS KAKENHI (Grant Numbers 19H05790, 20H00354 and 21H05233). B.A.F. acknowledges a Stanford Graduate Fellowship. Part of this work was performed at the Stanford Nano Shared Facilities (SNSF), supported by the National Science Foundation under award ECCS-2026822.   

\section{Author contribution}
B.A.F, J.Y., and B.E.F. designed and conducted the scanning SET experiments. B.A.F. and B.E.F. designed and conducted the dual gate capacitance experiments. T.D., Y.Z., and L.F. conducted theoretical calculations. B.A.F. fabricated the samples, with help from C.R.K. K.W. and T.T. provided hBN crystals. All authors participated in analysis of the data and writing of the manuscript.

\section{Competing interests}
The authors declare no competing interest. 


\end{document}


\title{Supplementary Material for:\\Tunable spin and valley excitations of correlated insulators in $\Gamma$-valley moir\'e bands}

\author{Benjamin A. Foutty}
\thanks{These authors contributed equally}
\affiliation{Geballe Laboratory for Advanced Materials, Stanford, CA 94305, USA}
\affiliation{Department of Physics, Stanford University, Stanford, CA 94305, USA}

\author{Jiachen Yu}
\thanks{These authors contributed equally}
\affiliation{Geballe Laboratory for Advanced Materials, Stanford, CA 94305, USA}
\affiliation{Department of Applied Physics, Stanford University, Stanford, CA 94305, USA}

\author{Trithep Devakul}
\thanks{These authors contributed equally}
\affiliation{Department of Physics, Massachusetts Institute of Technology, Cambridge, Massachusetts 02139, USA}

\author{Carlos R. Kometter}
\affiliation{Geballe Laboratory for Advanced Materials, Stanford, CA 94305, USA}
\affiliation{Department of Physics, Stanford University, Stanford, CA 94305, USA}

\author{Yang Zhang}
\affiliation{Department of Physics, Massachusetts Institute of Technology, Cambridge, Massachusetts 02139, USA}

\author{Kenji Watanabe}
\affiliation{Research Center for Functional Materials, National Institute for Material Science, 1-1 Namiki, Tsukuba 305-0044, Japan}

\author{Takashi Taniguchi}
\affiliation{International Center for Materials Nanoarchitectonics, National Institute for Material Science, 1-1 Namiki, Tsukuba 305-0044, Japan}

\author{Liang Fu}
\affiliation{Department of Physics, Massachusetts Institute of Technology, Cambridge, Massachusetts 02139, USA}

\author{Benjamin E. Feldman}
\email{bef@stanford.edu}

\affiliation{Geballe Laboratory for Advanced Materials, Stanford, CA 94305, USA}
\affiliation{Department of Physics, Stanford University, Stanford, CA 94305, USA}
\affiliation{Stanford Institute for Materials and Energy Sciences, SLAC National Accelerator Laboratory, Menlo Park, CA 94025, USA}

\maketitle

\tableofcontents

\newpage

\section{1. Spatial variation measured with scanning SET}
\begin{figure}[h!]
    \renewcommand{\thefigure}{S\arabic{figure}}
    \centering
    \includegraphics[scale =1]{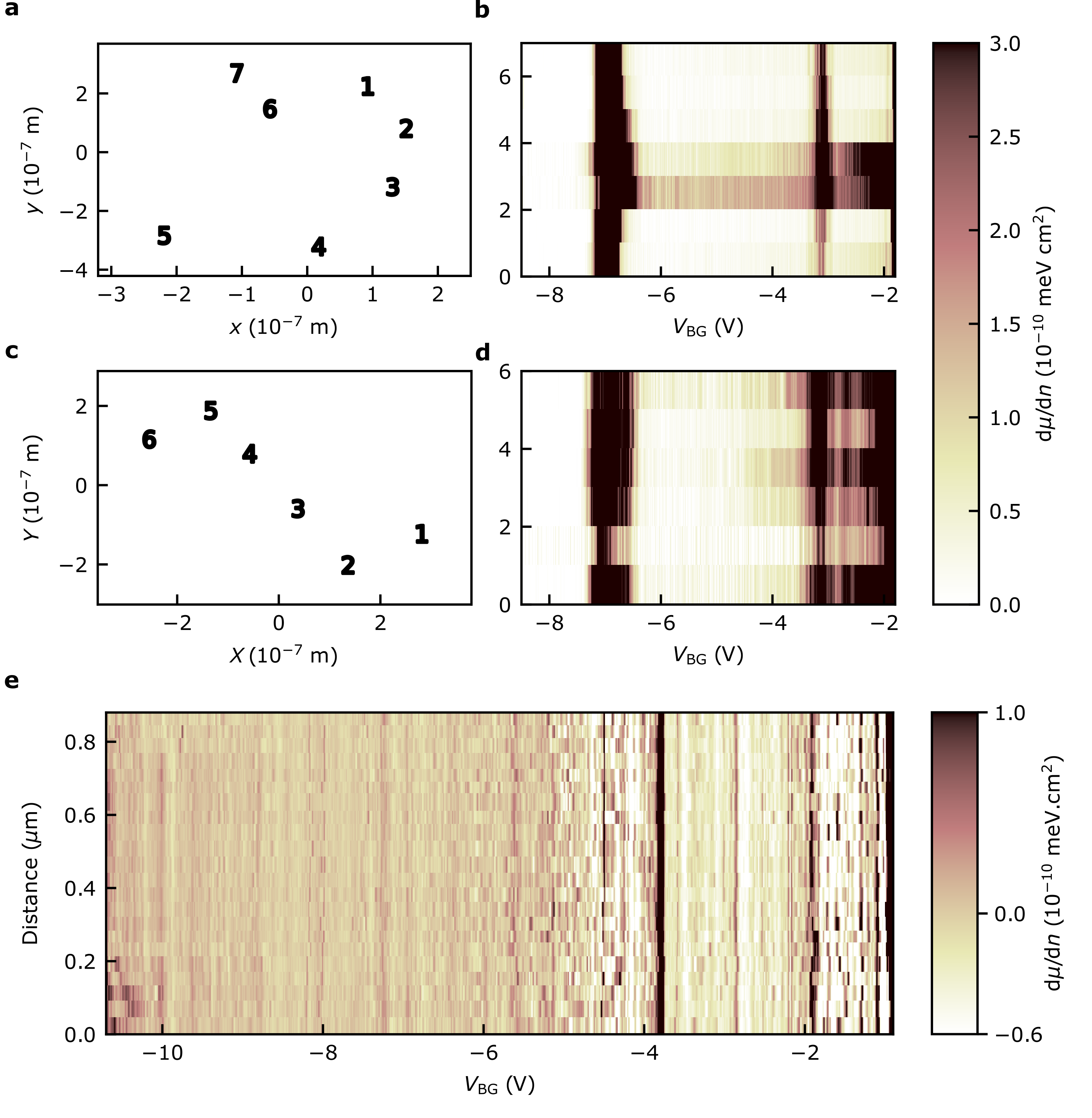}
    \caption{\textbf{Spatial dependence at $B = 11 \textrm{ T}$}. \textbf{a}, Measured locations in Area 1 of Sample S1. $x$ and $y$ label relative spatial coordinates. \textbf{b}, Inverse electronic compressibility d$\mu/$d$n$ as a function of back voltage $V_{\textrm{BG}}$ at the positions shown in \textbf{a}, labelled by index. \textbf{c}, Locations of measurements in a far-separated region of Sample S1. $X$ and $Y$ label spatial coordinates within this second region. \textbf{d}, d$\mu/$d$n$ at positions shown in \textbf{c}. \textbf{e}, Spatial variation of d$\mu$/d$n$ along a continuous linecut in Sample S2.} 
    \label{fig:Spatial}
\end{figure}

\begin{figure}[h]
    \renewcommand{\thefigure}{S\arabic{figure}}
    \centering
    \includegraphics[scale =1]{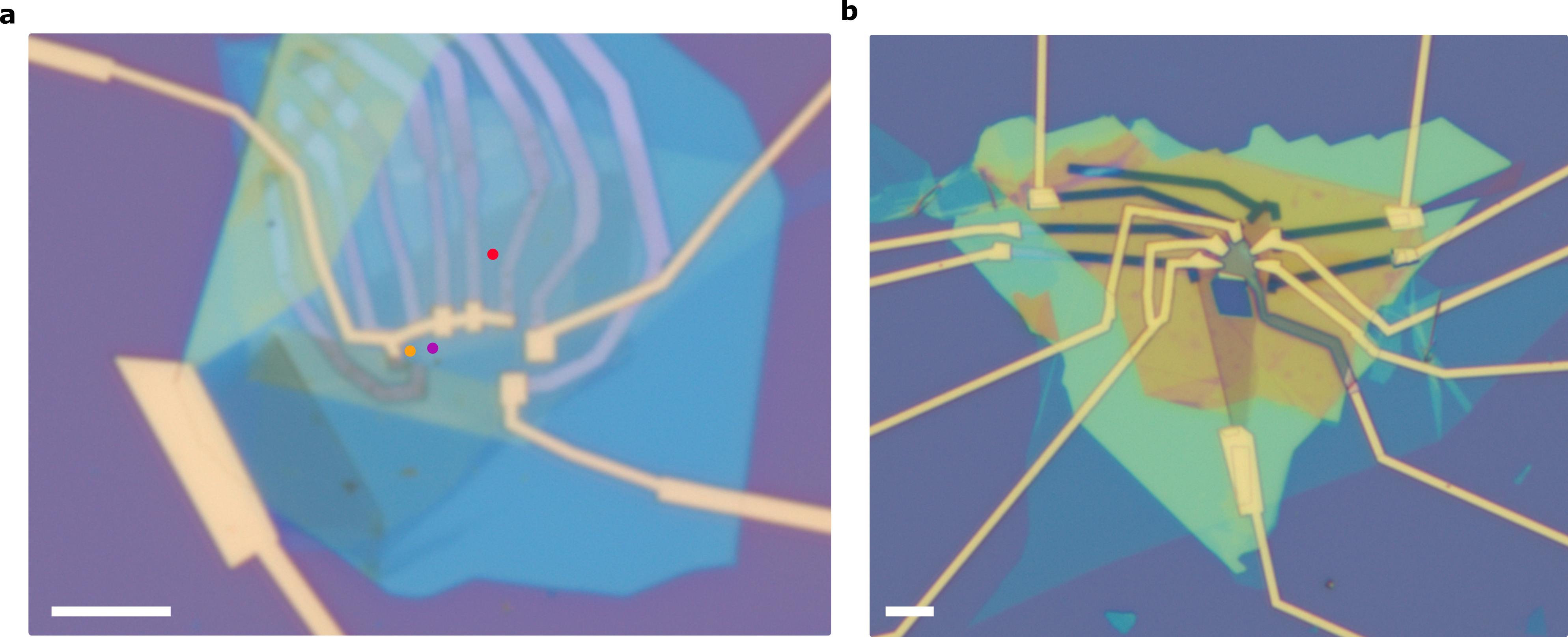}
    \caption{\textbf{Optical images of the samples}. \textbf{a}, Optical image of Sample S1. The measurements in the main text are all taken in the vicinity of the purple dot, while similar behavior was observed in the vicinity of the red dot (Fig. \ref{fig:Spatial}c-d) and measurements of the adjacent Bernal bilayer region were taken at the orange dot (Fig. \ref{fig:LLs}). \textbf{b}, Optical image of Sample C2. Scale bars in \textbf{a} and \textbf{b} are both 5 $\mu$m.} 
    \label{fig:DeviceImages} 
\end{figure}

In the main text, we focus on a single spatial location in each sample. In Fig. \ref{fig:Spatial}, we present spatial dependence of inverse electronic compressibility $d\mu/dn$ taken in each sample at magnetic field $B=11$ T, where the incompressible states are strongest. Here, $\mu$ is the chemical potential and $n$ is the carrier density, with the convention that increasing hole density is negative. Both devices are highly homogeneous over micron length scales, with little twist angle disorder in the regions examined. We estimate the total twist angle disorder by fitting the positions of the measured incompressible states and comparing their spatial dependence. In Sample S1, we compare the difference in gate voltage between the peak at filling factor $\nu = -1$, which occurs around back gate voltage $V_{\textrm{BG}} = -7$ V and the $\nu = -\frac{1}{3}$ peak, which occurs around $V_{\textrm{BG}} = -3.2$ V. This yields a twist angle that varies between $\theta = 3.38^\circ$ and $\theta = 3.45^\circ$ in Area 1, where the measurements in the main text were conducted (Fig. \ref{fig:Spatial}a-b, Fig. \ref{fig:DeviceImages}a). In a second region (Area 2) in Sample S1 that is about 4 $\mu$m away, the twist angle varies between $\theta = 3.42^\circ$ and $\theta = 3.61^\circ$ (Fig. \ref{fig:Spatial}c-d, Fig. \ref{fig:DeviceImages}a). In Sample S2, we fit the difference between the incompressible $\nu=-1$ gap at $V_{\textrm{BG}} = -3.95$ V and the incompressible peak at $V_{\textrm{BG}} = -3$ V, and estimate a twist angle that varies between $\theta = 2.55^\circ$ and $\theta = 2.65^\circ$ (Fig. \ref{fig:Spatial}e). We note that this peak may not be a charge-ordered state, but appears to occur at a roughly constant filling factor position and therefore can be used as a benchmark to estimate the twist angle disorder that behaves better than the band edge, which can vary due to underlying potential disorder in the sample. If we assume a constant band edge and just treat the position of the $\nu = -1$ gap in gate voltage as the only parameter, this leads to an even smaller estimate of twist angle disorder. Across both devices, we measure significantly less twist angle inhomogeneity than is typically observed across magic-angle twisted bilayer graphene devices \cite{uri_mapping_2020,zondiner_cascade_2020,yu_correlated_2022}. This difference likely results from the larger overall twist angles in these devices which limit excessive relaxation/reconstruction and/or increased stiffness due to the presence of more layers.

\begin{figure}[h!]
    \renewcommand{\thefigure}{S\arabic{figure}}
    \centering
    \includegraphics[scale =1]{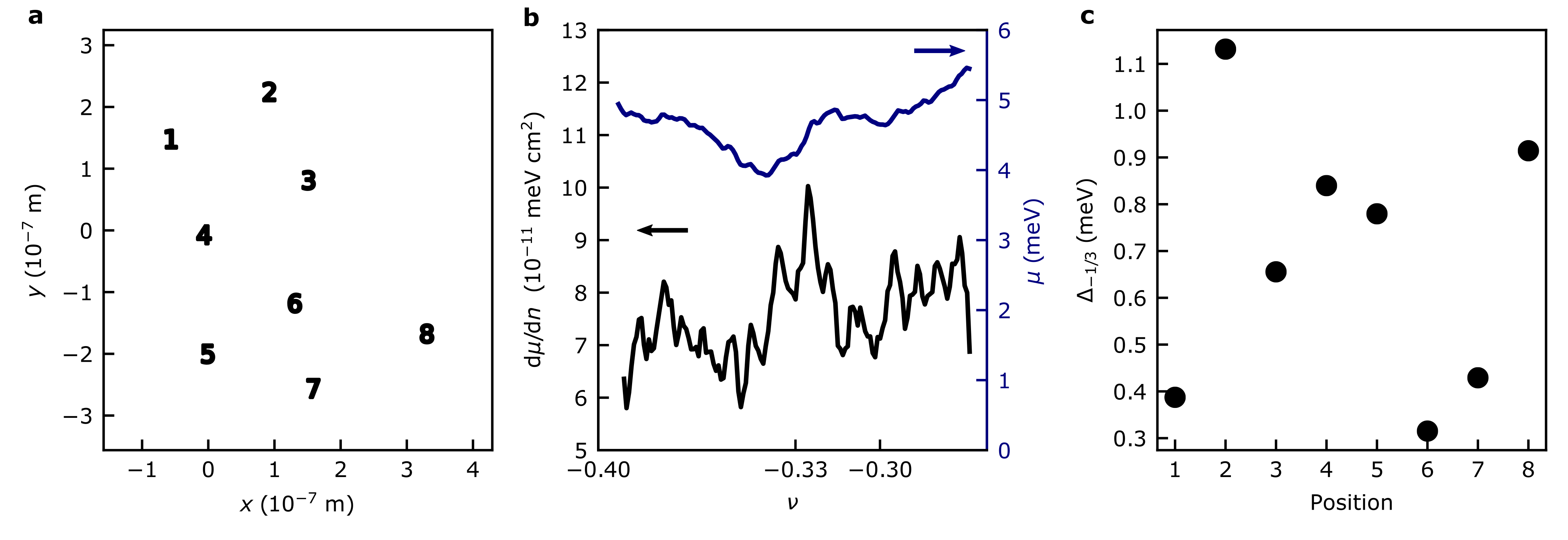}
    \caption{\textbf{Spatial dependence of the $\Delta_{-1/3}$ at $B=0$}. \textbf{a}, Spatial location of points plotted in panel \textbf{c}, relative to (0,0) as set in Fig. \ref{fig:Spatial}. \textbf{b}, Left, $\textrm{d}\mu/\textrm{d}n$ at $B=0$ around $\nu = -1/3$ and right, integrated chemical potential $\mu(n)$ subtracting out a flat background at $7\times 10^{-11}$ meV cm$^2$. The resulting gap $\Delta_{-1/3}$ is taken to be the difference in the local maximum and minimum in $\mu(n)$. \textbf{c}, Gap size $\Delta_{-1/3}$ at $B =0$ as a function of spatial position as enumerated in \textbf{a}.}
    \label{fig:13spatial}
\end{figure}

In Sample S1, the gaps at $\nu = -1$ are prominent at a variety of locations both at $B = 0$ and $B = 11$ T. However, the $\nu = -\frac{1}{3}$ gap is much closer to the noise floor of the measurement at $B = 0$. In Fig. \ref{fig:13spatial}, we present spatial dependence of this gap at $B=0$ across a 400 nm $\times$ 500 nm region of the sample. We measure a small but nonzero gap at $\nu = -\frac{1}{3}$ between 0.3 meV and 1.1 meV over  this area. The variability in gap size likely arises due to local variation in the underlying spatial disorder (from static charge disorder, small amounts of twist angle variability, and/or strain), which can appreciably affect the magnitude of the small gap at $B=0$.

\section{2. Gap extraction}
The gaps $\Delta_\nu$ presented in the main text correspond to the step in the chemical potential at filling factor $\nu$, found by taking the difference between the local maximum and local minimum in $\mu$ on either side of the gap. In the scanning single-electron transistor (SET) measurements, we use the DC measurement of $\mu(n)$ to extract charge gap sizes, rather than integrating the AC measurement of $d\mu/dn$ because the high resistance of the contacts and/or sample artificially enhances the a.c. signal of the incompressible peaks at high fields. 

\begin{figure}[h]
    \renewcommand{\thefigure}{S\arabic{figure}}
    \centering
    \includegraphics[scale =1]{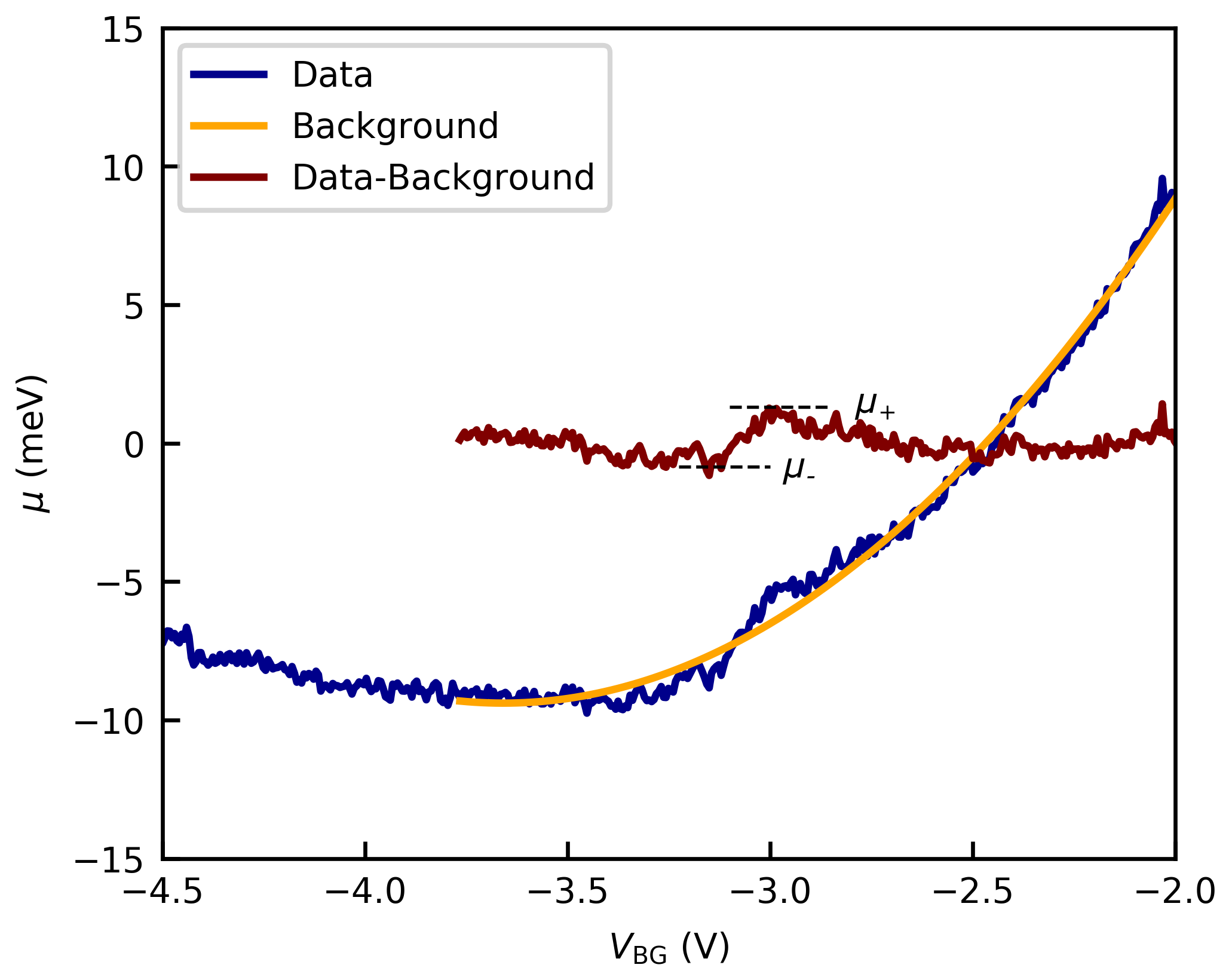}
    \caption{\textbf{Background fitting near $\nu=-1/3$}. Background subtraction around the $\nu=-1/3$ gap in Sample S1, measured at $B = 5$ T. Black dashed lines show local maximum and minimum values of $\mu_+$ and $\mu_-$ around the gap, the difference of which is taken to be $\Delta_{-1/3}$.}
    \label{fig:nu13}
\end{figure}

\begin{figure}[h]
    \renewcommand{\thefigure}{S\arabic{figure}}
    \centering
    \includegraphics[scale =1]{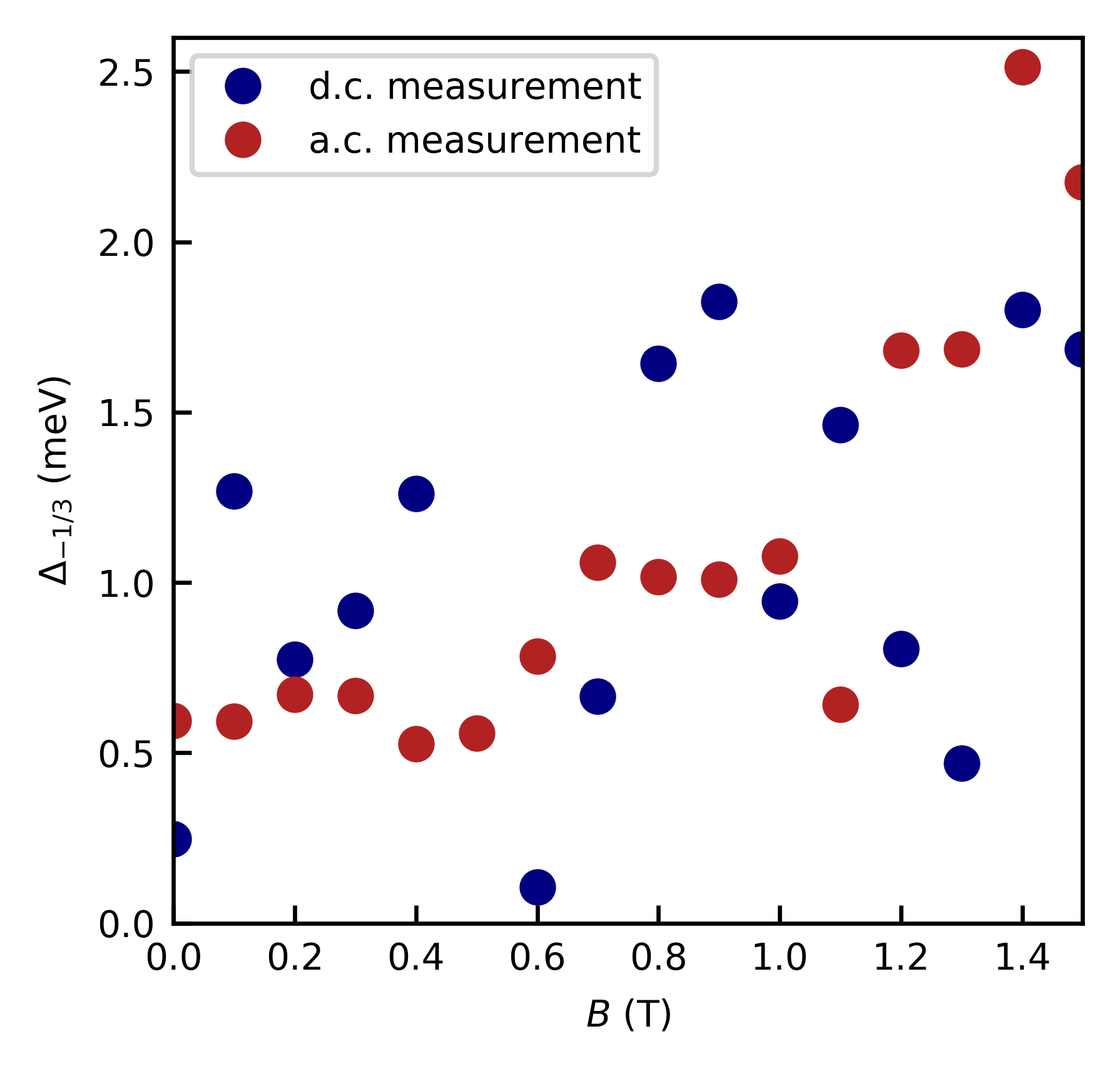}
    \caption{\textbf{Comparison of a.c. and d.c. measured gaps at $\nu=-1/3$ at low magnetic fields}. The a.c. measurement, which has overall lower noise, falls well within the scatter of the d.c. measured data. The two measurements are performed in the same location simultaneously, and the d.c. gaps are identical to those presented in Fig. 2c. Above B = 1.5 T, the a.c. measurement suffers from artificial enhancement, hence the focus on the d.c. measured gap in the main text.} 
    \label{fig:ACvDC}
\end{figure}

In Sample S1, the $\nu=-1/3$ gap appears as $\mu(n)$ begins to turn up near the mobility edge even in the d.c. measurement scheme due to high sample resistance at low densities. To accurately extract the gap, we fit a quadratic polynomial curve to the surrounding data and subtract this background before evaluating the gap size, avoiding an overestimation due to the background shape and finite width of the incompressible gap. Variation of the bounds of this fitted background do not affect the measured gap size nor its quantitative trend with $B$ (i.e., the effective $g$-factor). An example showing data at 5 T is plotted in Fig. \ref{fig:nu13}. After subtracting the quadratic background, the size of the d.c. measured gap agrees within the noise with the simultaneously a.c. measured gap at low fields (Fig. \ref{fig:ACvDC}), where a nearly density-independent background is straightforward to identify before integrating d$\mu$/d$n$. This is the only gap for which we perform any background subtraction; other gaps are well separated from the mobility edge.

In the global capacitance measurements, we only have direct access to $d\mu/dn$, rather than the chemical potential itself. To determine the charge gaps, we convert the capacitance signal to $d\mu/dn$ (see Supplementary Sec. 13 below) and then integrate over density to find $\mu(n)$. The charge gaps are then taken as the resulting step on this $\mu(n)$ curve.

\section{3. Continuum modeling}
We derive effective continuum models for twisted double bilayer WSe$_2$ (tdWSe$_2$) from first principles calculations at high symmetry stacking configurations.

The low energy physics of hole-doped monolayer WSe$_2$ consists of (local) quadratic band maxima at the $\Gamma$ and $\pm K$ points of the Brillouin zone.  
The $\Gamma$ valley band maximum is spin-degenerate, while the $\pm K$ band maxima are spin polarized with spin $s_z=\pm \frac{1}{2}$ due to spin-orbit coupling.
The structure of four-layer WSe$_2$ is such that each layer is rotated $180^\circ$ relative to its adjacent layers, so that interlayer tunneling is strongly suppressed in the $\pm K$ valley bands. 

Three different high-symmetry stacking configurations, MM$^\prime$, MX$^\prime$, and XX$^\prime$, as illustrated in the main text, are considered.  
First principles density functional theory (DFT) calculations for four-layer WSe$_2$, shown in Fig. 1e of the main text, reveal that the valence band maximum is at $\Gamma$ for all three stacking configurations.  
The energies of the top four valence bands at $\Gamma$, and the top two at $K$, are listed in Table~\ref{tab:gamma}.

\begin{table}
\renewcommand{\thetable}{S\arabic{table}}

\begin{tabular}{ c| c c c c c c}
 Band &  $E_\Gamma$ (MM$^\prime$) & $E_\Gamma$ (MX$^\prime$) &   $E_\Gamma$ (XX$^\prime$)  &  $E_K$ (MM$^\prime$) & $E_K$ (MX$^\prime$) &   $E_K$ (XX$^\prime$)\\ 
 \hline 
 1 & -4.5164 & -4.7022 & -4.6223 & -4.6523 & -4.8350 & -4.7064\\
 2 & -4.6412 & -4.8144 & -4.7011 & -4.6602 & -4.8569 & -4.7168\\
 3 & -4.9506 & -5.1330 & -5.0938 & -&- &-\\
 4 & -5.3199 & -5.5013 & -5.3622 & -& -&-
\end{tabular}
\caption{Energy (in eV) obtained from DFT at the $\Gamma$ and $K$ points, for each of the three high symmetry stacking regions MM$^\prime$, MX$^\prime$, and XX$^\prime$.  Note that each band is doubly degenerate due to inversion and time reversal symmetry.  }\label{tab:gamma}
\end{table}

We first consider an effective single-particle continuum model for the twisted system of the form
\begin{equation}
H_{\Gamma} = -\frac{|\mathbf{k}|^2}{2m_{\Gamma}} + V_{\Gamma}(\mathbf{r})
\end{equation}
where $m_{\Gamma}$ is the $\Gamma$ valley effective mass, and $V(\mathbf{r})$ describes a periodic potential with the moir\'e periodicity.
The spin degeneracy is implicit.
The potential $V(\mathbf{r})$ can be obtained by taking the first harmonic approximation, $V_{\Gamma}(\mathbf{r}) = 2 v \sum_{i=1,3,5}\cos (\mathbf{g}_i\cdot\mathbf{r}+\phi) + c$,
from which the three parameters, $v,c,\phi$, are chosen to match the energy of the first $\Gamma$ band in Table~\ref{tab:gamma} at the three positions $\mathbf{R}_{\mathrm{MM}^\prime},\mathbf{R}_{\mathrm{MX}^\prime}$, and $\mathbf{R}_{\mathrm{XX}^\prime}$. 
Here, $\mathbf{g}_i$ are moir\'e reciprocal lattice vectors defined $\mathbf{g}_i=\frac{4\pi}{\sqrt{3}a_M}(-\sin\frac{2\pi (i-1)}{6},\cos\frac{2\pi(i-1)}{6})$,
and  $\mathbf{R}_{\mathrm{MM}^\prime}=0$, $\mathbf{R}_{\mathrm{MX}^\prime}=\frac{a_M}{\sqrt{3}}(0,-1)$, and $\mathbf{R}_{\mathrm{XX}^\prime}=\frac{a_M}{\sqrt{3}}(0,1)$.
With this definition, $V(\mathbf{R}_{\mathrm{MM}^\prime})=6v\cos(\phi)+c$, $V(\mathbf{R}_{\mathrm{MX}^\prime})=6v\cos(\phi+2\pi/3)+c$, and $V(\mathbf{R}_{\mathrm{XX}^\prime})=6v\cos(\phi-2\pi/3)+c$.
We find the parameters $v=17.9$ meV, $\phi=0.44$ (radians), and $c=-4613.6$ meV.
We use the moir\'e period $a_M=3.32\ \mathrm{\AA}/(2\sin(3.4^\circ/2))\approx 56\ \mathrm{\AA}$, and effective mass $m_\Gamma=1.2 m_e$, where $m_e$ is the electron mass~\cite{movva_tunable_2018}.
The hole density obtained from filling the first moir\'e band of this continuum model is shown in Fig. 1f of the main text.

To incorporate the effect of a vertical displacement field, we now construct a model which includes all four layer degrees of freedom.
We assume the form for the model Hamiltonian at zero displacement field $H_{\Gamma} =  -\frac{|\mathbf{k}|^2}{2 m_\Gamma}\mathbb{I} + W_{\Gamma}(\mathbf{r})$ where $\mathbb{I}$ is the identity matrix and
\begin{equation}
W_{\Gamma}(\mathbf{r}) = 
 \begin{pmatrix} 
 V_{1\Gamma}(\mathbf{r}) & T_{1\Gamma}(\mathbf{r}) & 0 & 0 \\
T_{1\Gamma}(\mathbf{r}) &  V_{2\Gamma}(\mathbf{r})& T_{2\Gamma}(\mathbf{r}) & 0 \\
0 & T_{2\Gamma}(\mathbf{r}) & V_{2\Gamma}(\mathbf{r}) & T_{1\Gamma}(\mathbf{r}) \\
0 & 0 & T_{1\Gamma}(\mathbf{r}) & V_{1\Gamma}(\mathbf{r})
\end{pmatrix},
\end{equation}
in which all $T$, $V$, are periodic functions with the moir\'e periodicity, and are chosen
 such that the eigenvalues of ${W}_{\Gamma}(\mathbf{R}_\alpha)$
 with $\alpha=MM^\prime,MX^\prime,XX^\prime$, match with $E_{\Gamma}(\alpha)$ in Table~\ref{tab:gamma}.
To uniquely specify $T$, we have taken it to be real and positive.
At each $\mathbf{R}_\alpha$, there are four parameters $\{V_{1\Gamma}(\mathbf{R}_\alpha),V_{2\Gamma}(\mathbf{R}_\alpha),T_{1\Gamma}(\mathbf{R}_\alpha),T_{2\Gamma}(\mathbf{R}_\alpha)\}$, which can be chosen to fit the four eigenvalues $E_{\Gamma}$.  
From these, we can use the first harmonic approximation to obtain 
\begin{equation}
\begin{split}
V_{1\Gamma}(\mathbf{r})/\mathrm{meV} &=  -4799.5 + 16.4 \sum_{i=1,3,5}2\cos(\mathbf{g}_i\cdot\mathbf{r}+0.62)  \\
V_{2\Gamma}(\mathbf{r})/\mathrm{meV} &=  -5093.6 + 18.5 \sum_{i=1,3,5}2\cos(\mathbf{g}_i\cdot\mathbf{r}+0.47)  \\
T_{1\Gamma}(\mathbf{r})/\mathrm{meV} &=   218.6 + 1.2 \sum_{i=1,3,5}2\cos(\mathbf{g}_i\cdot\mathbf{r}+1.84) \\
T_{2\Gamma}(\mathbf{r})/\mathrm{meV} &=   220.3 -7.8 \sum_{i=1,3,5}2\cos(\mathbf{g}_i\cdot\mathbf{r}+2.18)  
\end{split}
\end{equation}
which specifies the continuum model.

\begin{figure}
    \renewcommand{\thefigure}{S\arabic{figure}}
\includegraphics[width=0.4\textwidth]{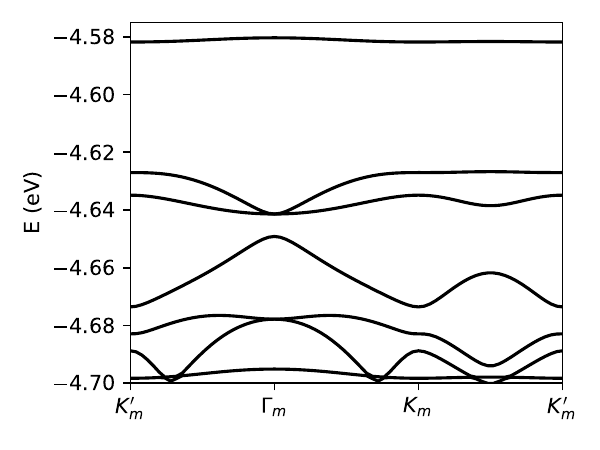}
\includegraphics[width=0.4\textwidth]{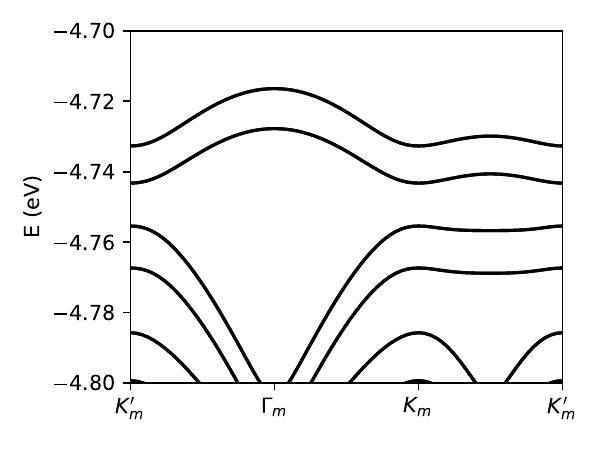}
\caption{\textbf{Moir\'e bands from the continuum model.} The $\Gamma$ (left) and $K$ (right) valley moir\'e bands obtained from the four layer continuum model.   }\label{SMFig:Bands}
\end{figure}

A similar construction can be made for the $K$ or $K^\prime$ valley bands.
We focus on the $K$ valley, which is related to the $K^\prime$ valley by time reversal.
The main difference from the $\Gamma$ bands is that interlayer tunneling is suppressed, and hence $H_K=-\frac{|\mathbf{k}|^2}{2m_K} + \mathrm{diag} ( V_{1K}(\mathbf{r}), V_{2K}(\mathbf{r}), V_{2K}(\mathbf{r}), V_{1K}(\mathbf{r}))$,
and $\mathbf{k}$ is now measured relative to the $K$ point.
The first two energies $E_K$ at each stacking region specify the potentials $V_{1K}$ and $V_{2K}$. However, there is now an additional ambiguity as the eigenvalues are invariant under interchanging $V_{1K}\leftrightarrow V_{2K}$.  
Additional information is needed to determine whether the first pair of bands should be associated with $V_{1K}$ (the outer two layers) or $V_{2K}$ (the inner two layers).
From direct inspection of the DFT wave function, we find that the first band has larger component on the outer two layers at all stacking regions.
Hence, $V_{1K}>V_{2K}$ and the potentials are fully specified in the first harmonic approximation:
\begin{equation}
\begin{split}
V_{1K}(\mathbf{r})/\mathrm{meV} &=  -4731.2 + 18.1 \sum_{i=1,3,5}2\cos(\mathbf{g}_i\cdot\mathbf{r}+0.75)  \\ 
V_{2K}(\mathbf{r})/\mathrm{meV} &=  -4744.6 + 19.5 \sum_{i=1,3,5}2\cos(\mathbf{g}_i\cdot\mathbf{r}+0.76)  
\end{split}
\end{equation}
We use the effective mass $m_K=0.5 m_e$~\cite{movva_tunable_2018}.
The moir\'e band structure for the $\Gamma$ and $K$ valley bands is shown in Fig~\ref{SMFig:Bands}.

To address the competition between $\Gamma$ and $K$ valley bands in a displacement field,
we model the displacement field by a layer-dependent potential difference $\Delta$ via the term $H_{\Delta} = -\Delta\;\mathrm{diag}(0,1,2,3)$.
At $\nu=0$, the layer potential is related to the displacement field $D$ by $\Delta = d D / \epsilon$, where $\epsilon$ is the relative out-of-plane dielectric constant and $d$ is the interlayer spacing.

\begin{figure}
    \renewcommand{\thefigure}{S\arabic{figure}}
\includegraphics[height=6cm]{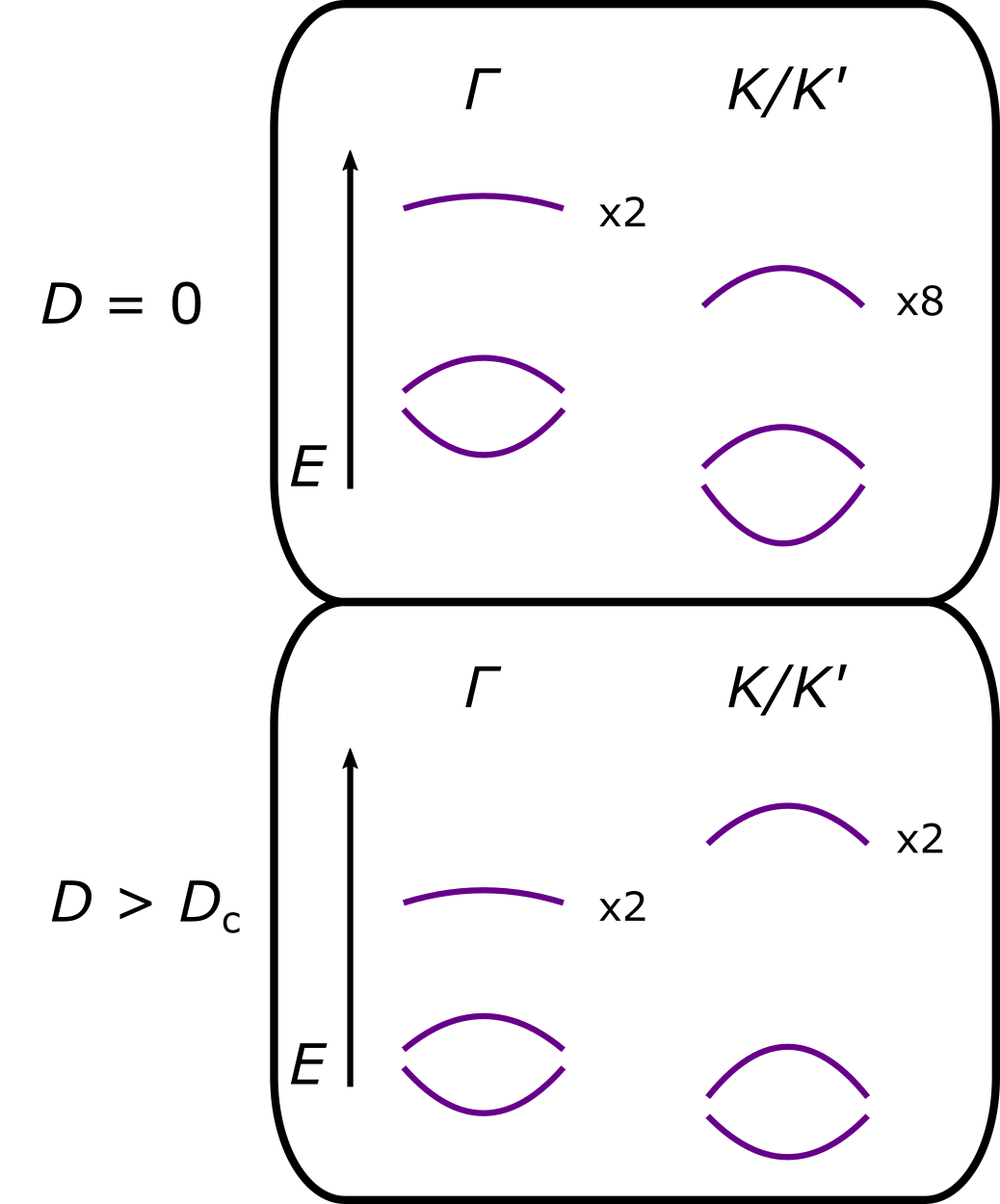}
\caption{\textbf{Schematic of single-particle moir\'e bands at low and high $D$.} At $D = 0$, the lowest energy single-particle moir\'e bands are two-fold (spin) degenerate at the Gamma point. The subsequent moir\'e band, localized at the $K / K^\prime$ points, is eight-fold degenerate (4x layer, 2x spin/valley). At $D > D_c$, the critical displacement field at which the $K / K^\prime$ moir\'e band is the lowest energy band, the lowest energy moir\'e band is completely localized on a single layer as favored by the displacement field with a two-fold degeneracy from spin/valley (up at $K$ / down at $K^\prime$, or vice versa depending on the direction of $D$). The next moir\'e band will be the same two-fold degenerate $\Gamma$ moir\'e band.}

\label{SMFig:MoireBandCartoon}
\end{figure}

We remark that the competition between $\Gamma$ and $K$ band edges from first principles is often delicate, and can depend sensitively on relaxed lattice structure and van der Waals density functional (see, for instance, Refs.~\cite{wu_topological_2019} and \cite{devakul_magic_2021}, in which different functionals result in different $\Gamma$,$K$ band orderings in both twisted homobilayer WSe$_2$ and MoTe$_2$).
In addition, interaction effects may also result in a renormalization of the band gap measured experimentally.
To compare quantitatively with experiments, we thus take as an additional tuning parameter a constant energy offset $\Delta_{ K}$ between $\Gamma$ and $K$ band edges, modifying the Hamiltonian as $H_{\Gamma}\rightarrow H_{\Gamma}-\Delta_{\Gamma K}/2$ and $H_K\rightarrow H_K + \Delta_{\Gamma K}/2$.
The experimentally measured gap is maximum at $D=0$, and decreases continuously and smoothly with non-zero $D$:
this is consistent with the scenario in which the state at $\nu=-1$ corresponds to filling a spin-polarized $\Gamma$ moir\'e band and the gap being to the first $K$ or $K^\prime$ moir\'e band.
We choose $\Delta_{\Gamma K}=0.1$ eV such that the gap between the first $\Gamma$ and $K$ bands roughly match the experimentally measured value at $D=0$.  
Once $\Delta_{\Gamma K}$ is fixed, the dependence of the gap on $\Delta$ follows from the continuum model, as shown in Fig. 3f of the main text. The closing, rather than opening of the gap as a function of $D$ is only consistent with a $\Gamma$ moir\'e band being the lowest energy hole band at $D=0$. The band degeneracies at each valley and their relative energies are schematically illustrated in Fig. \ref{SMFig:MoireBandCartoon}.

We note that this theoretical picture is also consistent with the lack of a gap observed at $\nu = -2$ in the experiment. Specifically, at low displacement fields, our calculations suggest the states filled beyond $\nu = -1$ are in a second, more dispersive band localized at the $K$-valley. At $\nu = -2$, this second band will be partially filled, and its larger bandwidth makes it less likely to form a correlated insulator.

\begin{figure}
    \renewcommand{\thefigure}{S\arabic{figure}}
\includegraphics[width=0.8\textwidth]{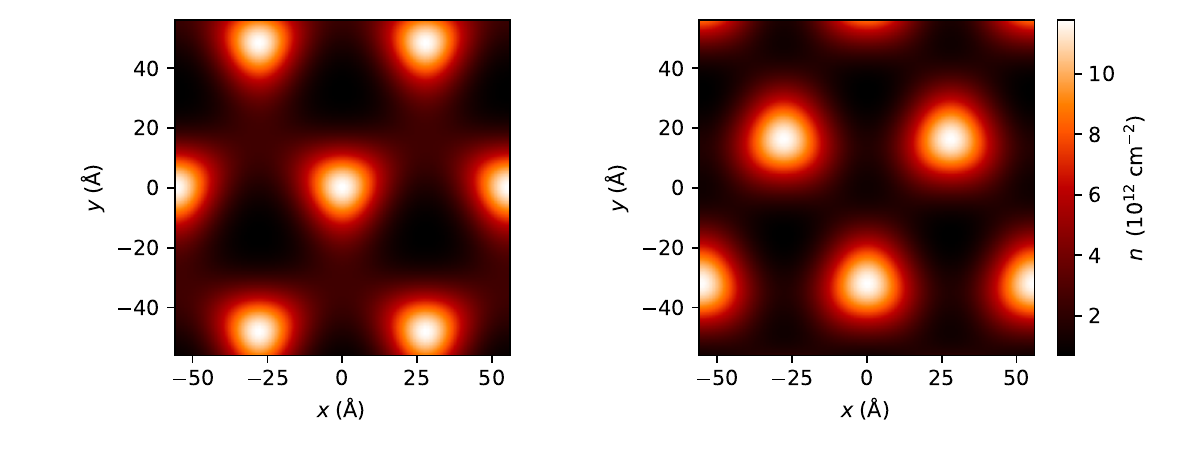}
\caption{\textbf{$K$-valley band density distribution.} Density distribution in the first $K$ valley band without (left) and with (right) the electrostatic potential from the filled $\Gamma$ band at $\nu=-1$, calculated from the continuum model.  
Without interactions, the charges are localized at MM$^\prime$, whereas including the electrostatic potential shifts them to XX$^\prime$.
}\label{SMFig:Hartree}
\end{figure}

To conclude this section, we comment on the nature of the lowest energy $K$ valley excitations at $\nu=-1$.  
In the absence of interactions, it is clear from the moir\'e potential that holes prefer to localize on the MM$^\prime$ sites for both $\Gamma$ and $K$ valley.
However, holes filling into the $K$ valley beyond $\nu=-1$, will feel an effective potential resulting from electrostatic repulsion with the filled $\Gamma$ band.
This effect can be captured by a Hartree potential in the continuum model,
$
V_H(\mathbf{r})=\int d\mathbf{r} V_C(\mathbf{r}-\mathbf{r}^\prime)\langle n^{(\Gamma)}(\mathbf{r}^\prime)\rangle
$
where $\langle n^{(\Gamma)}(\mathbf{r}^\prime)\rangle$ is the density distribution of the filled first $\Gamma$ band (obtained from the non-interacting continuum model), and $V_C(\mathbf{r})$ is the Coulomb potential.
The continuum model description of the first $K$ band now includes $V_H(\mathbf{r})$ in addition to the moir\'e potential.
We further simplify by taking $\mathbf{r}$ to be purely in the 2D plane and use a vertically displaced Coulomb potential $V_C(\mathbf{r}) = e^2/(4\pi\epsilon\sqrt{\mathbf{r}^2+d^2})$, where $\epsilon=3.5\epsilon_0$ and $d=7$ \AA.
Figure~\ref{SMFig:Hartree} shows the hole density of the first $K$ valley band with and without the Hartree term.
As can be seen, holes in the $K$ valley prefer to localize in the XX$^\prime$ regions once electrostatic repulsion from the filled $\Gamma$ band is taken into account, resulting in an effective honeycomb lattice description near $\nu=-1$.

\section{4. Spin $g$-factors}
To quantitatively compare the experimental gaps and their increase in a magnetic field to our theoretical model, we must consider the Zeeman contribution to the effective $g$-factors of the various excitations. Our measurement of a gap at filling $\nu$, $\Delta_{\nu}\equiv \Delta\mu$ is equivalent to the particle-hole excitation energy, given by $E[N_\nu + 1] + E[N_\nu - 1] - 2 E[N_\nu]$, where $E[N]$ is the ground state energy at particle number $N$, and $N_\nu$ denotes the particle number at filling $\nu$. First, we consider the case at high magnetic fields (i.e., the right side of Fig. 2i in the main text), where the ground state is a spin-polarized insulator with charge transfer energy gap $\Delta_{\textrm{CT}}$, the high-density excitation is an additional interstitial hole with aligned spin, and the low-density excitation is the absence of a hole. Then, we can write down the energies as follows, which results in a gap that is independent of magnetic field: 
$$E[N_\nu] = -\frac{1}{2}g_sN_\nu B,\quad E[N_\nu+1] =\Delta_{\textrm{CT}} - \frac{1}{2} g_s (N_\nu+1)B,\quad E[N_\nu-1] = -\frac{1}{2}g_s (N_\nu-1)B$$
$$\Delta_\nu = E[N_\nu + 1] + E[N_\nu - 1] - 2 E[N_\nu] = \Delta_{\textrm{CT}}-\frac{1}{2}g_sB + \frac{1}{2}g_s B = \Delta_{\textrm{CT}}$$

At intermediate fields (i.e., the left side of Fig. 2i in the main text), where the only change is that the hole excitation is the formation of a spin polaron, we instead have
$$E[N_\nu] = -\frac{1}{2}g_sN_\nu B,\quad E[N_\nu+1] = \Delta_{\textrm{CT}}-E_b - \frac{1}{2} g_s (N_\nu+1)B,\quad E[N_\nu-1] = -\frac{1}{2}g_s (N_\nu-2)B+\frac{1}{2} g_s B$$
$$\Delta_\nu = E[N_\nu + 1] + E[N_\nu - 1] - 2 E[N_\nu] = \Delta_{\textrm{CT}}-E_b -\frac{1}{2}g_sB + \frac{3}{2}g_s B = \Delta_{\textrm{CT}}-E_b + g_s B$$
where $E_b>0$ is the spin polaron binding energy.
Given that the bare spin $g$-factor $g_s = 2$, the measured effective $g$-factor of our gapped states would be 2 in the absence of orbital/Berry curvature and exchange effects.

As mentioned in the main text, orbital and Berry curvature differences between different moir\'e bands (at $K$ and $\Gamma$) can lead to additional contributions to $g^*$ at $\nu = -1$. The momentum-resolved orbital magnetization $m(k)$ can vary widely across the the moir\'e Brillouin zone (mBZ) \cite{grover_chern_2022}. Theoretical calculation of $m(k)$ requires calculation of the Berry curvature on the single-particle moir\'e bands and can depend sensitively on shape of the moir\'e bands, and therefore the twist angle \cite{grover_chern_2022,he_giant_2020,devakul_magic_2021}. Because our probe is only sensitive to the lowest energy excitation, the measured gap will depend sensitively on the extremal value of $m(k)$ across the mBZ, which can lead to different effective $g$-factors between samples with different twist angles, and may contribute to the slightly different $g$-factors measured at $\nu = -1$ in Samples S1 and S2.

\section{5. Alternative excitations at $\nu = -\frac{1}{3}$}
\begin{figure}[h]
    \renewcommand{\thefigure}{S\arabic{figure}}
    \centering
    \includegraphics[scale =1]{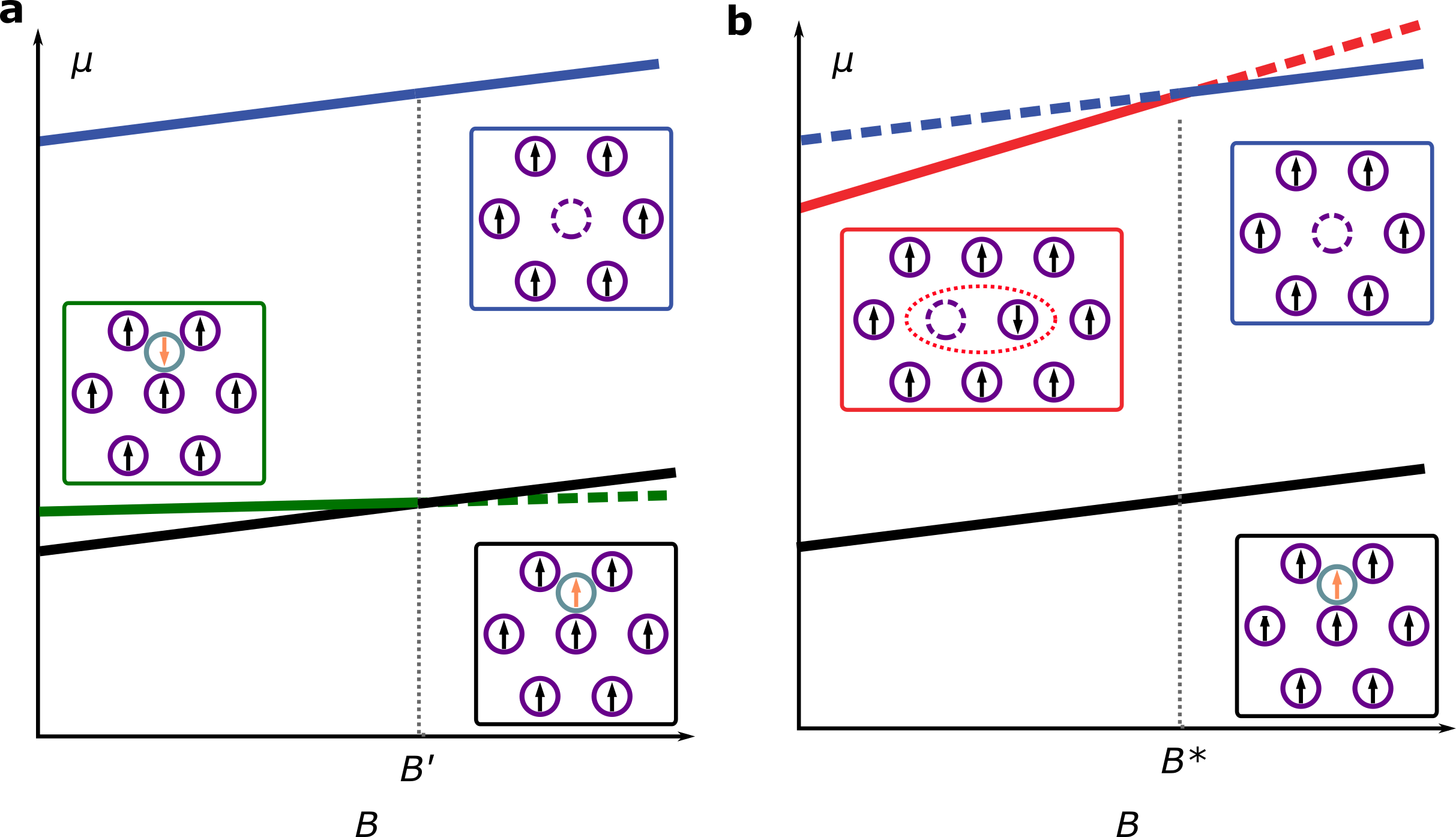}
    \caption{\textbf{Schematic of magnetic field dependence of possible excitations in the considered scenarios. }\textbf{a}, Analogue of Fig. 2i for the scenario detailed in Supplementary Sec. 5, in which the excitation below the gap changes from an anti-aligned to aligned hole at $B = B^\prime$, while the excitation above the gap is the removal of a hole at all fields. We show the chemical potential $\mu$ as a function of $B$ for various excitations of the (spin-polarized) correlated insulating ground state. The blue line and inset show an excitation above the gap in which a single hole is removed. The green line and inset, the favored excitation below the gap at low magnetic fields, show the addition of a spin anti-aligned hole at an interstitial site. The black line and inset show the addition of an aligned spin at an interstitial site, which is the favored excitation below the gap at high magnetic fields. \textbf{b}, Fig. 2i in the main text, reproduced here for ease of comparison.} 
    \label{fig:SuppFig_2ialt} 
\end{figure}

In the main text, we mention an alternative explanation of the saturating gap at $\nu = -\frac{1}{3}$, namely that below $B = 7$ T, exchange coupling causes the lowest energy additional hole to be anti-aligned with the spin-polarized ground state, until Zeeman coupling overcomes this barrier and the additional hole is aligned with the rest of the spins. In this section, we expand on that picture and discuss why it is not supported by our data and calculations. In this picture, the change in behavior around $B=7$ T occurs on the excitations below the gap (i.e., $\mu_-$ in the main text). In Fig. \ref{fig:SuppFig_2ialt}a, we show the expected chemical potentials for each excitation (above and below the gap) in this scenario, in an analagous way to Fig. 2i in the main text (which is reproduced as \ref{fig:SuppFig_2ialt}b for ease of comparison). If the changing spin character of the additional hole excitation were driving the saturating gap, then we would expect the discontinuity in the slope of the chemical potential on the band edge to occur at $\mu_-$, rather than as we observe at $\mu_+$.  Additionally, the energy scale of $J$ that our data would predict in this case does not appear to be consistent with other microscopic parameters of the problem. To lowest order, we could estimate exchange coupling $J$ as $\frac{1}{3}g^* \mu_B B^\prime$ where $g^*$ is the effective $g$-factor that we measure for the gap, $\mu_B$ is the Bohr magneton, $B^\prime = 7$ T is where the behavior of the gap changes, and the factor of $\frac{1}{3}$ is due to the fact that an added hole couples to three nearest neighbors. This yields $J \approx 0.6$ meV, which is quite large compared with other parameters in the problem. For example, if we take $J \sim 4t^2/U$ where $t$ is the hopping and $U$ is the onsite potential, then we would estimate $J \approx 0.01$ meV, as $t \approx 0.2$ meV from the bandwidth of the lowest moir\'e band from the continuum model while $U \approx 100$ meV due to the strength of Coulomb interactions \cite{zhang_moire_2020}.

The two scenarios of excitations considered in the main text operate off of the understanding of the insulators as spin-polarized throughout most of the measured range of magnetic fields. One possibility to consider as a potential explanation for the behavior of the gap at $\nu = -\frac{1}{3}$ is that rather than a change to the excitations at $B = 6$ T, the spin character of the gapped state itself changes, e.g. by becoming spin polarized above $B = 6$ T but being unpolarized below that field. We rule out this possibility as unlikely both due to theoretical considerations --– the next-neighbor spin-spin coupling should be very weak --- and experimental ones –- if the gapped state were not polarized by the magnetic field, we would not expect the gap to linearly increase in a magnetic field as measured in Fig. 2c. 

Further, the scenarios considered in the main text focus broadly on how the magnetic field couples to the spinful gapped states and excitations through Zeeman coupling. However, there are additional non-Zeeman orbital mechanisms by which the magnetic field can couple into the system.  We have considered two such possibilities: i) orbital effects within a moiré site can become important when the magnetic length is comparable to the size of the Wannier orbitals, leading to a change in the effective hopping/interaction strengths, ii) orbital effects can lead to a variation in the energy of excitations via formation of Hofstadter/LL bands. However, these effects should be almost negligible for realistic parameters: for our maximal magnetic field $B = 11$ T, the magnetic length $l_B \approx 8$ nm (which is already larger than the moir\'e period), and the slope of the gap, were it to come from a purely orbital mechanism, suggests an unrealistically small effective mass $m\approx 0.2 m_e$ for the hole excitations about the $\nu = -\frac{1}{3}$ state, and further cannot explain the discontinuity in the slope.  Additionally, we do not observe any prominent Hofstadter or LL gaps, which would be expected if such orbital effects were dominant.  While we cannot rule out the existence of more exotic mechanisms that we are not aware of, these considerations suggest that the role of orbital effects is minor in this system.  Nevertheless, further measurements in a tilted magnetic field could help clarify the relative strengths of orbital and Zeeman coupling.

Our measurements of the correlated insulating gaps provide evidence for spin polaron excitations, and more recent theory predicts a finite range of parameter space in which a metallic state involving such quasiparticles will be stabilized \cite{zhang_pseudogap_2022}. This compressible phase should have some novel physical properties which the measurements in the present manuscript are not sensitive to. For example, tunneling measurements should find that this state has a single particle gap (for tunneling in electrons) despite the fact that it is compressible. Additionally, there should be a simultaneous plateau in the magnetization throughout this phase. These measurements would provide complementary confirmation of novel itinerant quasiparticle excitations, and would be excellent directions for future experiments.

\section{6. Extraction of chemical potential $\mu_+$ and $\mu_-$}
The SET tip directly measures changes in chemical potential of the sample by tracking a particular Coulomb blockade oscillation of the SET while the sample density is tuned by a gate. In order to evaluate whether the discontinuity in the slope of the gap at $\nu = -\frac{1}{3}$ comes from a change on the particle or hole side of the gap, we need to evaluate the behavior of the chemical potential on either side of the gap as a function of magnetic field, which is the ``slow" axis in our measurement. One approach is to pick a particular point (say the band edge, or some location far from the gap) and choose this as a reference point at which $\mu$ is set to zero at every magnetic field. However, this can give different behavior depending on the choice of reference point, especially when the overall behavior of $\mu(n)$ changes smoothly with magnetic field. For example, in Fig. \ref{fig:Fig4}, it is apparent that the chemical potential at the band edge is controlled by the magnitude of negative compressibility; if the minimum point is chosen to be the ``reference" point where $\mu = 0$ for each value of $B$, then the trends in both $\mu_+$ and $\mu_-$ will mostly be determined by the change in the negative compressibility, rather than identifying the behavior in the vicinity of the gap itself.

To avoid such artifacts, we take advantage of the fact that the gap at $\nu = -\frac{1}{3}$ is small compared to the background chemical potential. We can fit a smooth polynomial background to the chemical potential on either side of the gap at $\nu = -\frac{1}{3}$ and subtract this background, as in Fig. \ref{fig:nu13}. Then the value of the chemical potential on either side of the gap is measured relative to the background chemical potential of filling the lowest energy moir\'e band between 0 and 1. The precise values of $\mu_+$ and $\mu_-$ and can vary slightly with details of the fit, but the fact that the $\textrm{d}\mu_+/\textrm{d}B$ changes at $B=7$ T while $\textrm{d}\mu_-/\textrm{d}B$ does not is consistent. We note that this essentially is a statement of the evolution of the chemical potential relative to the background filling of the lowest moir\'e band.

\begin{figure}[h]
    \renewcommand{\thefigure}{S\arabic{figure}}
    \centering
    \includegraphics[scale =1]{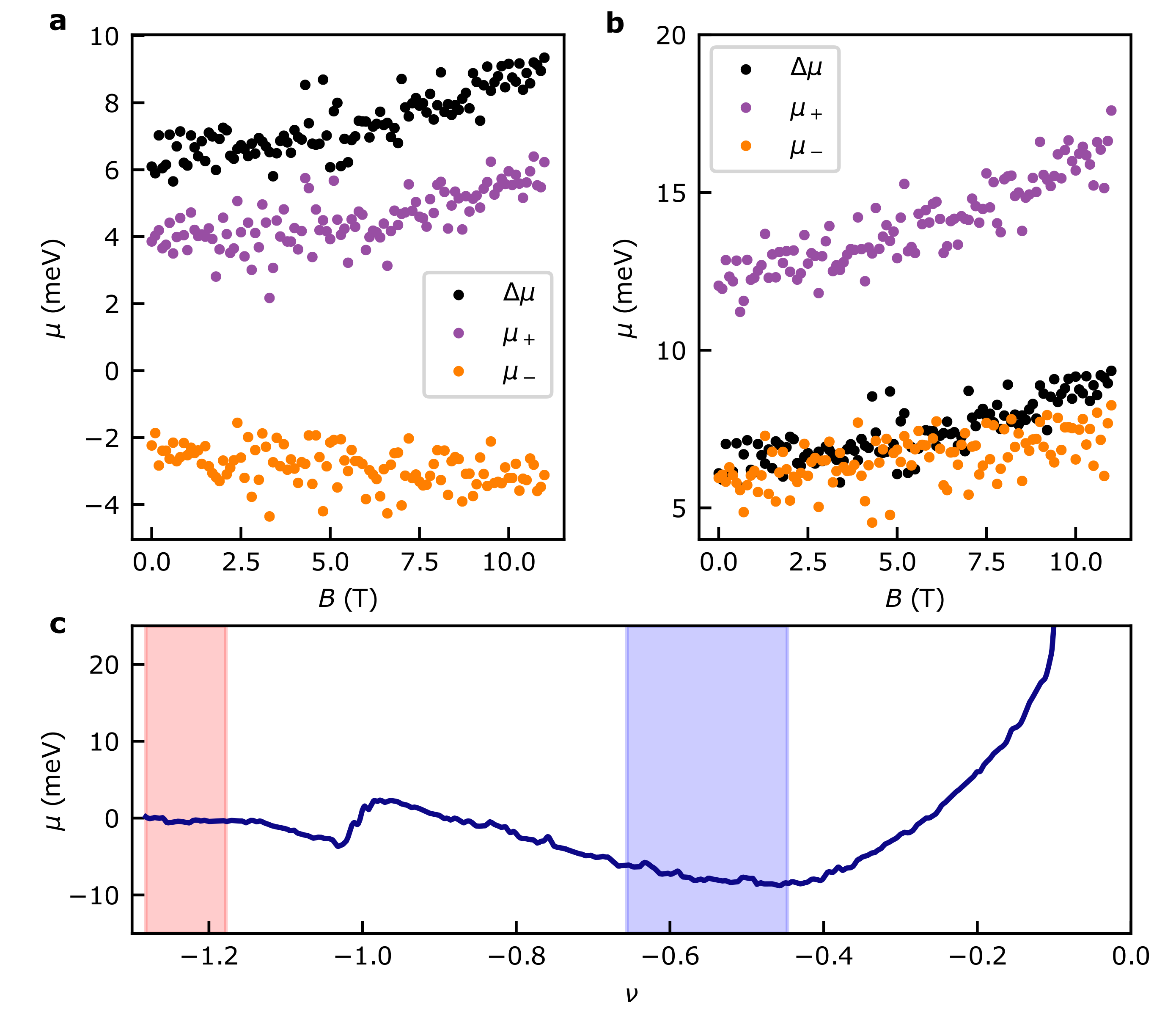}
    \caption{\textbf{Behavior of chemical potential around $\nu = -1$}. \textbf{a}, Chemical potential on the left  ($\mu_-$) and right ($\mu_+$) sides of $\nu = -1$ taking the reference $\mu$ at high densities around $\nu = -1.25$. \textbf{b}, Same as \textbf{a}, but with the reference $\mu$ set to intermediate densities around $\nu = -0.5$. \textbf{c}, $\mu(n)$ measured at $B=0$. The average $\mu$ within the red shaded box is set to zero for the data in \textbf{a}, while the average $\mu$ within the blue shaded box is set to zero for the data in \textbf{b}.} 
    \label{fig:nu1mu} 
\end{figure}

At $\nu = -1$, the behavior of $\mu(n)$ in both samples is dramatically different on either side of the gap, preventing an equivalent type of subtraction. Focusing on Sample S1, in Fig. \ref{fig:nu1mu} we show extracted chemical potential on either side of the $\nu = -1$ gap with two different choices of reference chemical potential. In Fig. \ref{fig:nu1mu}a, we set the reference chemical potential to $\nu = -1.25$. Practically, this means that we take the average measured chemical potential in a certain region (red shaded box in Fig. \ref{fig:nu1mu}c) and subtract it from the measurement at each field such that it becomes the ``zero" point of the chemical potential, before then reading off $\mu_+$ and $\mu_-$ on either side of the gap. In Fig. \ref{fig:nu1mu}b, we perform the same analysis but with the reference point taken near the middle of the first moir\'e band (blue shaded box in Fig. \ref{fig:nu1mu}c). In both of these cases, $\mu_-$ does not change very much; most of the gap dependence thus relates to the change of $\mu_+$. Given that we choose partially filled moir\'e bands as references, this suggests that the dependence of $\mu_-$ on magnetic field is quite close to the overall change in chemical potential of the bands themselves, while $\mu_+$ varies strongly relative to this. This qualitatively agrees with the theoretical picture described in the main text, as we would expect that at all experimentally accessible fields, the lowest-energy excitation at $\mu_-$ would be adding another spin-aligned hole, whose dependence on $B$ would be identical to filling the spin-polarized bands. The excitation at $\mu_+$ on the other hand, would be due to the formation of spin-polarons, which should have a different (stronger) dependence on magnetic field than the surrounding moir\'e bands (Fig. 2i in the main text). Therefore, our data also suggest spin polaron excitation are relevant at $\nu = -1$, but the smoking gun signature of a change in gap size and chemical potential evolution with magnetic field are not accessible in our measurement setup.

\section{7. Negative compressibility from exchange interactions}

\begin{figure}[h]
    \renewcommand{\thefigure}{S\arabic{figure}}
    \centering
    \includegraphics[scale=1.0]{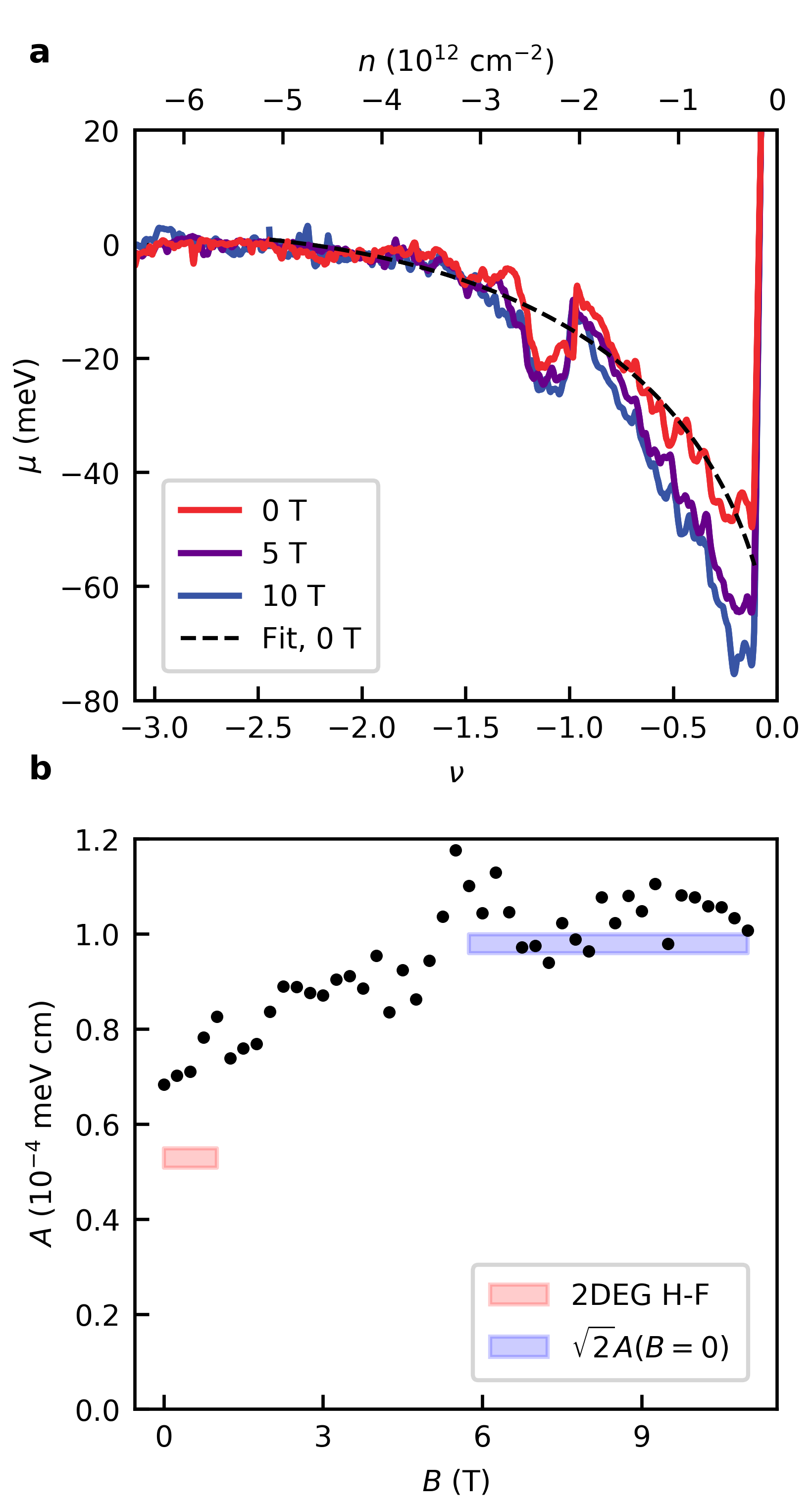}
    \caption{\textbf{Negative compressibility from exchange interactions}. \textbf{a}, $\mu(n)$ in Sample S2 at $B = 0, 5, 10$ T. Dashed line fit is to $\mu(n) = \frac{\pi\hbar^2}{m^*}n - A\sqrt{n}$ where $m^*$ and $A$ are free parameters. We set $\mu = 0$ at high hole density for each trace. \textbf{b}, Magnitude of the coefficient $A$ from the fit as a function of $B$. The red shaded box marks the $B=0$ prediction from a Hartree-Fock model of a 2D electron gas where the height shows uncertainty in dielectric environment. The blue shaded box marks the predicted growth in exchange due to full spin polarization, namely $\sqrt{2}\times A(B=0 \textrm{ T})$.} 
    \label{fig:Fig4}
\end{figure}

Here, we present evidence of strong interaction effects even away from the charge-ordered states. In Sample S2, we observe negative compressibility over a wide range of hole density, with $\textrm{d}\mu/\textrm{d}n$ becoming increasingly negative as hole density decreases (Fig. \ref{fig:Fig4}a). Negative compressibility is also ubiquitous in Sample S1 (Fig. 2b in the main text), but does not extend as close to the band edge due to a smooth, positive background signal for $|n| < 1.5\times 10^{12}$ cm$^{-2}$ (see below). We attribute this difference at low densities to varying TMD quality, possibly due to different commercial sources for the two samples, and focus on Sample S2 first.

Capacitance enhancements (equivalent to negative $\textrm{d}\mu/\textrm{d}n$) have been reported in TMD moir\'e heterostructures when the sample-gate distance $d$ is comparable to or less than the moir\'e length scale $L_M \approx a/\theta$ where $a$ is the lattice constant of WSe$_2$  \cite{li_charge-order-enhanced_2021,skinner_anomalously_2010}. In our devices, sample-gate coupling is small as $d/L_M \approx 4-5$, and much of the parameter space where we measure negative $\textrm{d}\mu/\textrm{d}n$ is in the limit of $nd^2 \gg 1$. In this regime, negative $\textrm{d}\mu/\textrm{d}n$ is generally attributed to strong Coulomb interactions \cite{eisenstein_negative_1992,eisenstein_compressibility_1994,li_very_2011,zondiner_cascade_2020,yang_experimental_2021}. We can understand the magnitude of negative compressibility by comparing to the Hartree-Fock (HF) expression for the chemical potential of a 2DEG: \begin{equation*}
    \mu(n) = \frac{\pi\hbar^2 n}{m^*} - \left(\frac{8}{\pi}\right)^{\frac{1}{2}} \left(\frac{e^2}{4\pi\epsilon}\right) \sqrt{n}
\end{equation*} These two oppositely signed terms describe the density of states and exchange contributions to the chemical potential and will always lead to negative $\textrm{d}\mu/\textrm{d}n$ at sufficiently low density \cite{eisenstein_negative_1992,eisenstein_compressibility_1994}. However, the low kinetic energy due to flat electronic bands in tdWSe$_2$ causes the negative contribution from exchange interactions to dominate $\mu(n)$ to much higher hole densities than in more dispersive electronic systems. This provides experimental evidence that the underlying moir\'e bands that we are filling in this density range are quite flat, consistent with our DFT calculations. At zero field, the extracted coefficient from a fit to this form (black dashed curve, Fig. \ref{fig:Fig4}a, see Supplementary Sec. 10 for fitting details) matches the order of magnitude of predictions from HF exchange based on the hBN dielectric environment (red shaded box, Fig. \ref{fig:Fig4}b). This coefficient, which correlates with the magnitude of the negative compressibility, grows monotonically before saturating around 7 T (Fig. \ref{fig:Fig4}b). While the exact value of this coefficient is sensitive to details of the fit, the order of magnitude and relative growth in a magnetic field are consistent (Supplementary Sec. 10, below). This increase is due to polarization of the compressible states being filled, as spin polarization will lead to an increase in the exchange interaction. Again in analogy to the 2DEG HF model, this increase saturates when the metallic phase is fully polarized, with a total increase by a factor $\sqrt{2}$, consistent with our observations (blue shaded box, Fig. \ref{fig:Fig4}b) \cite{tanatar_ground_1989}. This suggests that the magnetic ordering is paramagnetic through much of the density range, outside of the correlated insulator gaps.  

\begin{figure}[h]
    \renewcommand{\thefigure}{S\arabic{figure}}
    \centering
    \includegraphics[scale =1]{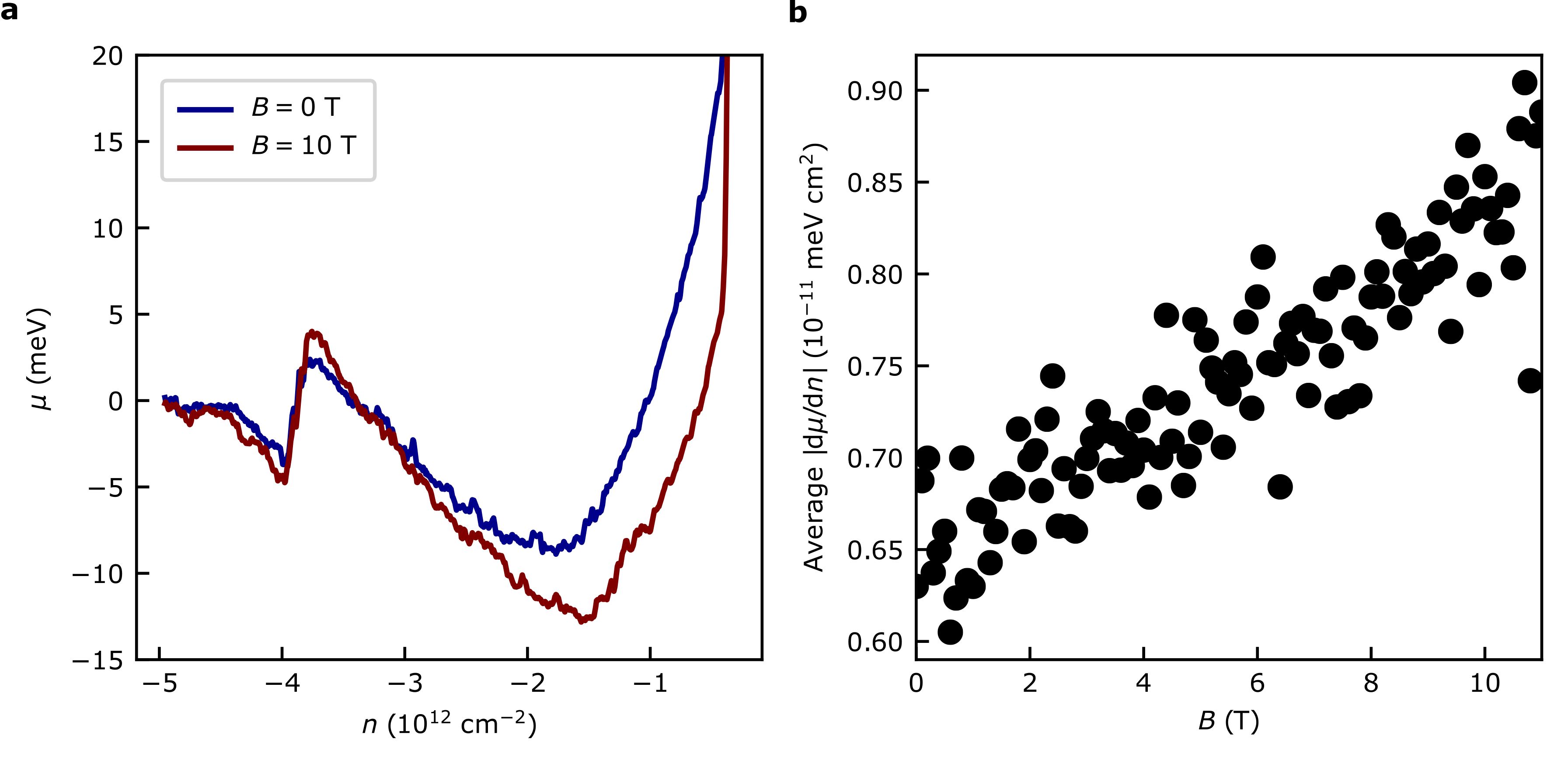}
    \caption{\textbf{Negative d$\mu/$d$n$ in Sample S1}. \textbf{a}, $\mu(n)$ as measured in S1 across the entire density range. \textbf{b}, Average value of $|\textrm{d}\mu/\textrm{d}n|$ between $n = -3.8\times 10^{12}$ cm$^{-2}$ and $n = -2.5\times 10^{12}$ cm$^{-2}$.} 
    \label{fig:S1DC} 
\end{figure}

Though we focus on Sample S2 in the discussion above, we also observe negative compressibility over large regions of hole doping in Sample S1, before an upturn at low hole density (Fig. \ref{fig:S1DC}). The magnitude of negative compressibility, as estimated from the average $|\textrm{d}\mu/\textrm{d}n|$ between $\nu = -.6$ and $\nu = -0.95$ also grows in a magnetic field, though no clear saturation is observed. 

\section{8. Fitting negative compressibility $\mu(n)$}
\begin{figure}[h!]
    \renewcommand{\thefigure}{S\arabic{figure}}
    \centering
    \includegraphics[scale =1]{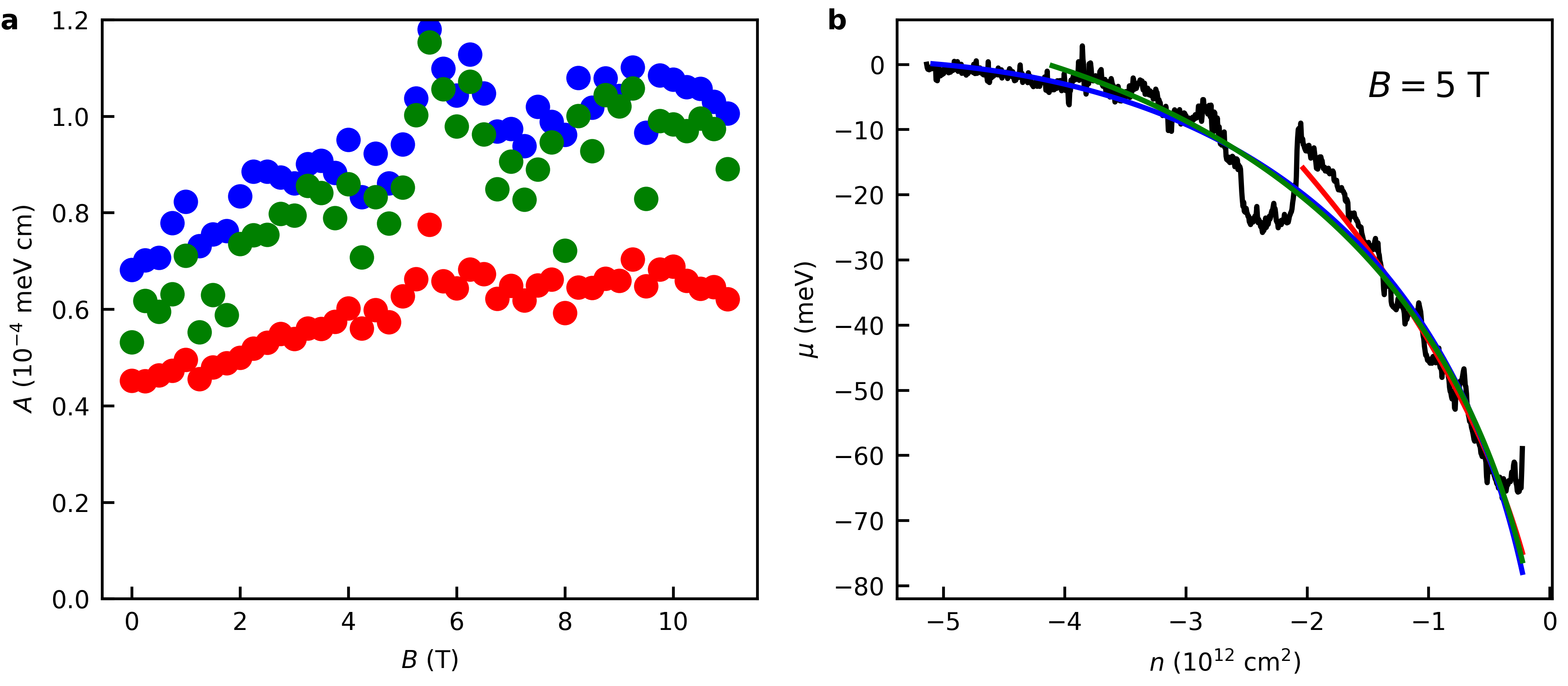}
    \caption{\textbf{Dependence of coefficient $A$ on fit parameters}. \textbf{a}, Extracted coefficient $A$ from a fit to $\mu(n) = \frac{\pi\hbar^2}{m^*} n - A\sqrt{n}$ where $m^*$ and $A$ are free parameters. Blue, green, and red data correspond to fitting the entire curve out to $n = -5\times 10^{12}$ cm$^{-2}$, $n = -4\times 10^{12}$ cm$^{-2}$, and $n = -2\times 10^{12}$ cm$^{-2}$ ($\nu = -1$), respectively. \textbf{b}, Measured $\mu(n)$ at $B = 5T$ (black) overlaid with fitted lines corresponding to those described in panel \textbf{a}.}
    \label{fig:Fitting1}
\end{figure}

Here we expand on the details of the fit to $\mu(n)$ in Sample S2 shown in Fig. \ref{fig:Fig4}. While the order of magnitude of the coefficient $A$ is robust to variation of fitting parameters/schemes, as is its increase by a factor of $\sqrt{2}$ from $B = 0$ to high fields, the precise value of $A$ is somewhat sensitive to details of the fitting procedure. In Fig. \ref{fig:Fitting1}, we present the extracted $A$ from 3 different fits, changing the range of density included in the fit. The blue data is that which is presented in the main text, because fitting over the widest possible density range seems to provide the best fit to the data. However, given the complicated details of the moir\'e bands, it is not clear which of these is most physically justifiable. Fitting across such a large density range means that we include filling over multiple moir\'e bands, each of which will have a varying effective density of states as they are filled. Additionally, it involves fitting across the charge-ordered gap at $\nu = -1$, around which the effective density $n$ may be better approximated by $|n-n_{\nu}|$, where $n_\nu$ is the charge density at moir\'e band filling $\nu$. It is apparent in these fits that between $n = -2\times 10^{12}\ \textrm{cm}^{-2}$ and $n = -2.5\times 10^{12}\ \textrm{cm}^{-2}$, just beyond $\nu = -1$, the behavior of $\mu(n)$ varies sharply compared to the background. For the red line, which fits to just the lowest moir\'e band between $\nu = 0$ and $\nu = -1$, the fit has no linear component. In the blue and green fits in Fig. \ref{fig:Fitting1}b, the size of the linear term is $\frac{\pi\hbar^2}{m^*} \approx 1.25\times 10^{-11}\ \textrm{meV/cm}^{-2}$ and $\frac{\pi\hbar^2}{m^*} \approx 1.75\times 10^{-11}\ \textrm{meV/cm}^{-2}$, respectively. Both of these terms would suggest effective masses much smaller than expected for moir\'e bands ($m^* = 0.1 -0.15 m_e$), which is inconsistent with the size of the Landau level gaps we observe simultaneously (Fig. \ref{fig:LLs}). Because these fits are over the entirety of multiple moir\'e bands, the effective mass is not necessarily expected to remain constant across the fit.

The analogy to the Hartree-Fock (HF) model in tdWSe$_2$ is a simplification, but it nonetheless remains qualitatively accurate, as detailed below. Specifically, the free electron gas calculation does not incorporate electron-electron correlations nor does it account for the moir\'e potential, which affects the kinetic energy. To determine how these details alter predictions for the chemical potential, we apply DFT to directly study the continuum model of interacting electrons in the periodic moir\'e potential, as developed in \cite{zhang_density_2020}. We use the local spin density functional (LSD) approximation, obtained from fixed node diffusion quantum Monte Carlo simulations of a two dimensional electron gas. By construction, the LSD approximation is exact for homogeneous systems. In practice, DFT with the LSD approximation provides a better description of moir\'e systems with smooth potential compared to the inhomogeneous systems: atoms, molecules and natural solids. In moiré materials with a smooth potential landscape and small bandwidth ($\sim$ 10 meV), the strongly correlated physics happens within the lowest few bands, and we can safely choose the energy cutoff as 10 eV. This leads to using $31 \times 31$ plane wave components, independent of the moiré wavelength or number of atoms in one unit cell.

\begin{figure}[h!]
    \renewcommand{\thefigure}{S\arabic{figure}}
    \centering
    \includegraphics[height=8cm]{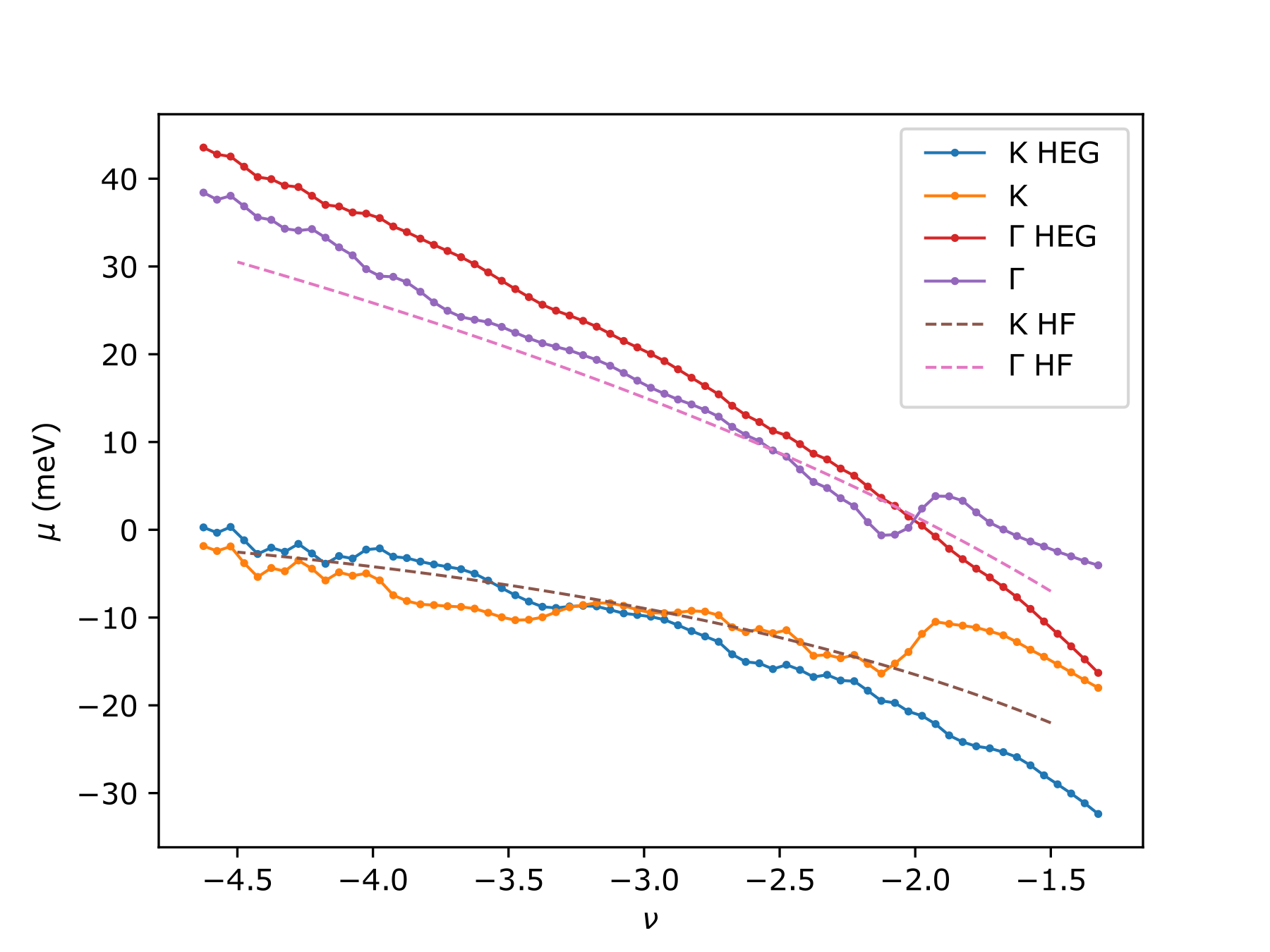}
    \caption{ \textbf{Moir\'e DFT calculations of chemical potential}. $\mu(\nu)$ calculated using moir\'e DFT for moir\'e bands at both $K$ and $\Gamma$ valleys. The red and blue curves, labeled HEG for homogeneous electron gas, show the limit where the moiré potential is set to zero, while the orange and purple curves take realistic values of the moir\'e potential, in line with DFT  calculations for tdWSe$_2$. The vertical offsets of these curves are arbitrary. 
}
    \label{fig:MoireDFTFitting}
\end{figure}

Because the experimental data spans a wide range of moir\'e fillings, it interpolates between filling $\Gamma$ and $K$ moir\'e bands at low and high fillings factors, respectively. In Fig. \ref{fig:MoireDFTFitting}, we show the resulting moir\'e  DFT calculations of $\mu(\nu)$ for both the $\Gamma$ and $K$ moir\'e bands. We note that this calculation is only performed for higher filling factors because it becomes unreliable at low filling, due to the possibility of magnetic and/or charge-ordered states. Due to this constraint, the $K$-valley calculations are more relevant to the corresponding portion of the experimental data, but we present both for full transparency. For each valley, we also present the HF prediction (dashed line) as a comparison as well as the homogeneous electron gas (HEG) prediction from the moir\'e DFT, which is the limit where the moir\'e potential is set to zero, but the correlations are calculated via DFT. All three curves are qualitatively similar, overlapping over broad ranges of density despite small quantitative differences at some fillings. The agreement between these theoretical calculations justifies applying the simpler HF expression to fit the negative compressibility over the full density range. The most prominent discrepancies arise near moir\'e filling factors (see for example, around $\nu = -2$), where gaps appear in the moir\'e DFT calculations but not in the HF model. This is similar to the experimental data, where there is sharp behavior (both the gap and subsequent sharper negative compressibility) between $\nu = -1.4$ and $\nu = -1$ in Fig. \ref{fig:Fig4} which is not captured by the smooth HF fit.

\section{9. Capacitance data from Sample C2}
\begin{figure}[h]
    \renewcommand{\thefigure}{S\arabic{figure}}
    \centering
    \includegraphics[scale =1]{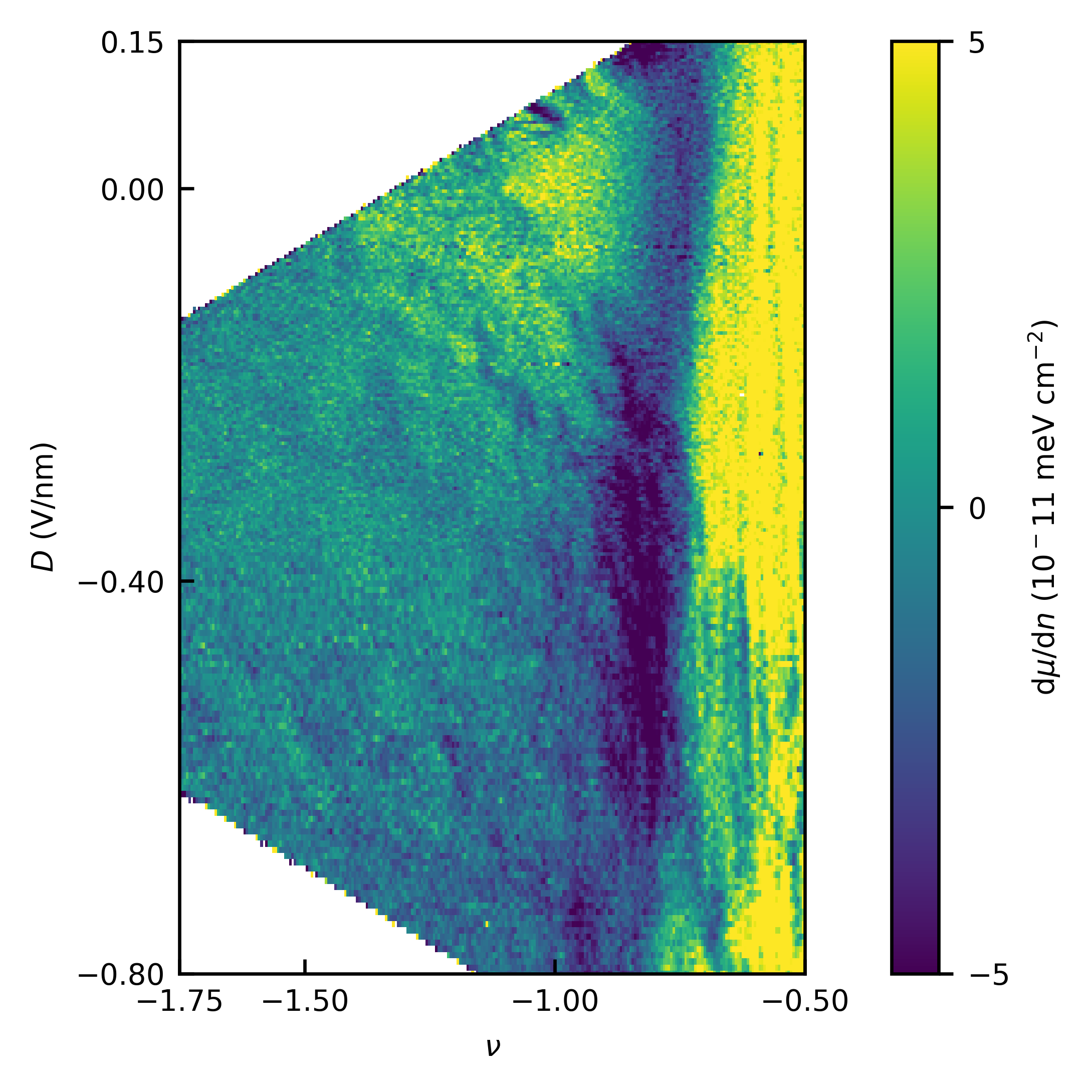}
    \caption{\textbf{Displacement field dependence in a second device}. Measured d$\mu$/d$n$ in Sample C2 as a function of displacement field $D$ and moir\'e filling factor $\nu$.} 
    \label{fig:C2dudn}
\end{figure}

In Fig. \ref{fig:C2dudn}, we present d$\mu$/d$n$ of Sample C2 as a function of $D$ and $\nu$. The data are extracted from ``top-gate capacitance" in which we apply a 10 mV a.c. excitation at 337 Hz to the top gate and measure the response on the HEMT gate, which is electrically connected to the sample contacts. The data were taken at $T = $ 4.2 K. The range that we can access on the lower (upper) left side are limited by top (bottom) gate dielectrics. 

\section{10. Estimating $D$ in SET measurements}
In general, the effective layer potential difference in a singly-gated two-dimensional (2D) heterostructure device is a difficult problem that depends in part on details of the material being gated \cite{kou_electron-hole_2014,mccann_asymmetry_2006}. Even the usual definition of an applied electric displacement field, which takes the average of the displacement field on either side of the gated 2D material, is an approximation and not necessarily directly convertible to microscopic parameters such as layer potential differences. Comparing our scanning SET measurements, in which the samples have no top gate, to measurements with a controlled displacement field is therefore nontrivial. However, a lowest order estimate can be made by approximating the set-up as a dual gated device where the SET tip is treated as an effective ``top gate" held at the same voltage as the sample. This voltage constraint is accurate because the relative sample-tip voltage is held by d.c. feedback in our experiment to minimize tip-induced doping. Additionally, the tip is held $\sim 50-100$ nm away in vacuum, such that the effective ``top-gate" capacitance will be quite small compared to that of the back gate, and any small offsets in voltage will not provide very much doping. We note that this is distinct from what has been reported in recent STM measurements, in which the close tip-sample distance requires the displacement field to be approximated by the voltage difference between tip and back gate \cite{kerelsky_moireless_2021}. Because the SET tip is far-separated and senses a spatial area underneath the tip that convolves both tip size and distance, any voltage separation between tip and sample leads to (small) non-uniform doping of the sample, resulting in broadening features rather than reliably doping the sample, which limits our ability to change the effective displacement field. 

Using this approximation, the effective displacement field can be written as $D_{\textrm{eff}}= \frac{1}{2\epsilon_0}(0 - C_b (V_b-V_0)) = -\frac{1}{2\epsilon_0} en$. Here, $C_b$ is the back gate capacitance, $V_b$ is the back gate voltage, $V_0$ is offset of the density due to intrinsic sample doping, and we've used that $C_b(V_b-V_0)/e = n$. Applying this expression to relevant densities in our samples, in Sample S1 at $\nu = -1$, $D_{\textrm{eff}}= 0.35$ V/nm, and in sample S2, $D_{\textrm{eff}} = 0.18$ V/nm; this nonzero displacement field explains the reduction in gap size relative to the $D=0$ value we measure in dual gated devices. We note that for Sample S1, this estimated $D$ is near and slightly above the critical displacement field at which we stop seeing insulating behavior at $\nu = -1$ in dual gated capacitance devices, though the precise value varies slightly across the two devices measured. However, the fact that we measure a gap at all at $\nu =-1$ in the singly-gated devices is strongly suggestive that the lowest energy moir\'e bands being filled are still at $\Gamma$, and suggest that the estimated $D_{\textrm{eff}}$ should be taken as a rough approximation. Additionally, global measurements in our dual gated devices inevitably average over disorder, which may lead to lower overall gap sizes relative to local sensing with the scanning SET. This will artificially decrease the critical displacement field at which the $\nu = -1$ gap vanishes in the dual gated devices. 

\section{11. Twist angle and band edge determination}
\begin{figure}[h!]
    \renewcommand{\thefigure}{S\arabic{figure}}
    \centering
    \includegraphics[scale =1]{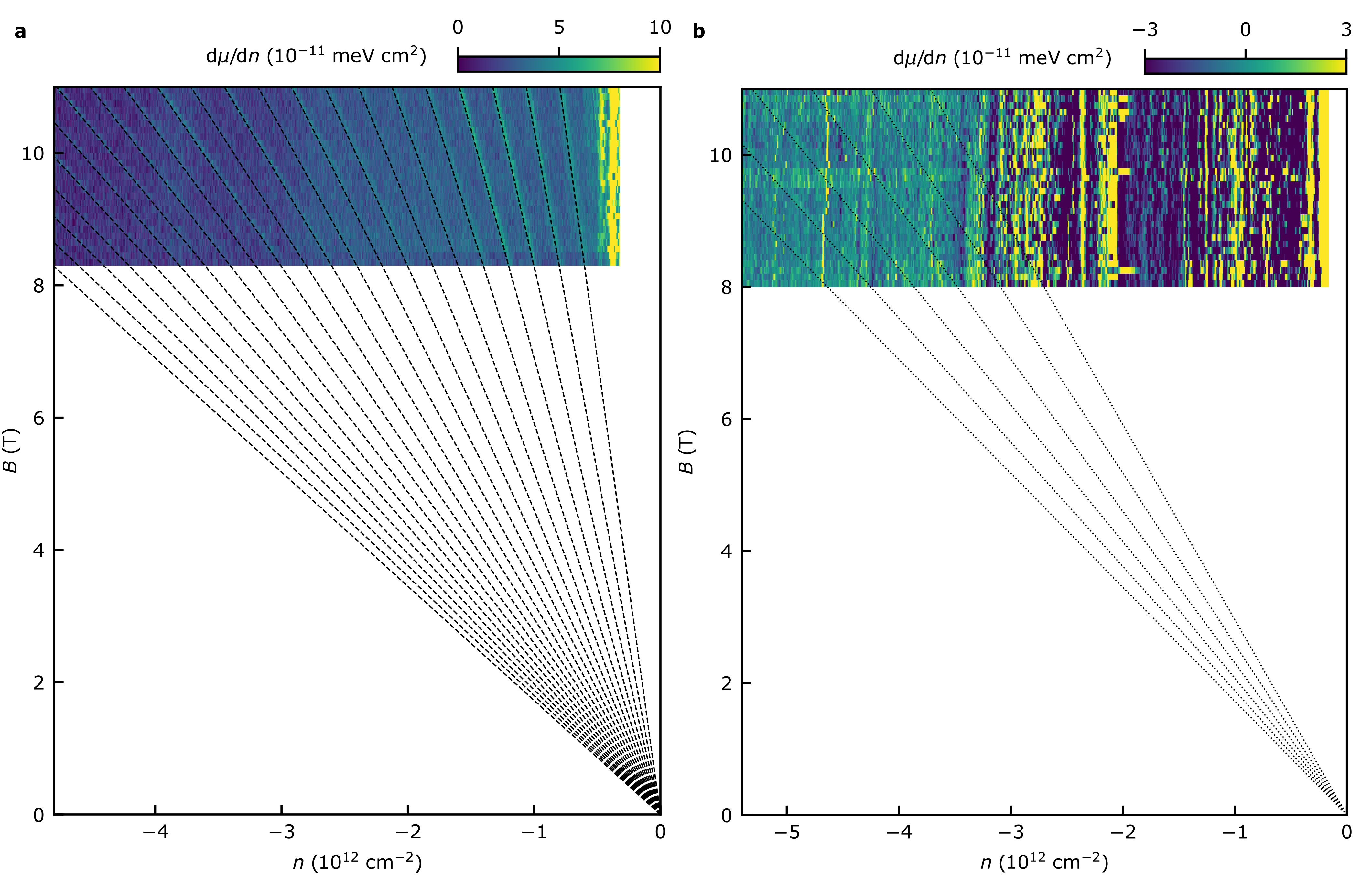}
    \caption{\textbf{Landau levels and band edge determination}. \textbf{a}, Landau levels observed in a Bernal bilayer region in Sample S1. Dotted lines are filling factor $\nu = 3,4,\ldots,24,25$. No clear Landau levels were observed in the tdWSe$_2$ region of the device. \textbf{b}, Landau levels in tdWSe$_2$ beyond $\nu = -1$ in Sample S2. These show a two-fold degeneracy; dotted lines are filling factor $\nu = 14,16,\ldots,24$.} 
    \label{fig:LLs}
\end{figure}

In order to quote accurate densities, we need to determine the location of the band edge in gate voltage, which is set by both local (global) intrinsic doping in the sample for SET (dual gated) measurements and by work function differences between gates and sample \cite{gustafsson_ambipolar_2018}. For both samples measured with the SET, the gate voltages at which we observe Landau level gaps allows for unambiguous determination of the band edge, which also matches well to the mobility edge (Fig. \ref{fig:LLs}). In Sample S1, we do not observe clear Landau levels in the tdWSe$_2$ region of the device (Fig. 2a), which may reflect the large effective mass of the moir\'e bands and/or the proximity to charge ordered states at the low fillings we measure (see below). In comparison, Landau levels are present in an adjacent Bernal bilayer WSe$_2$ region of Sample S1, from one of the ``halves" of the flake that makes up the device. Assuming the entire flake has similar intrinsic doping, this sets the gate voltage offset of the band edge in the twisted double bilayer region of the device as well. In Sample S2, we observe two-fold degenerate landau levels between $\nu = -1$ and $\nu = -3$ near the noise floor of our measurement, with gap sizes of order $200-300\ \mu$V at $B=11$ T. This corresponds to a filling factor range which we were unable to explore in Sample S1.

Here, we expand our discussion on the absence of Landau levels at low densities in both samples. The overall static disorder in these samples is consistent with previous reports. From Fig. \ref{fig:LLs}a, we measure a full-width at half-maximum of $4\times 10^{10}$ cm$^{-2}$ for the incompressible peaks of quantum Hall states in the adjacent Bernal bilayer region in Sample S1. This estimate of disorder is similar to quoted values in previous high-quality measurements of monolayer WSe$_2$ \cite{gustafsson_ambipolar_2018}. Additionally, we measure very little spatial variation in the twisted double bilayer regions of both devices (Supplementary Sec. 1). Thus, we rule out spatially inhomogeneous twist angle as a reason for the absence of Landau levels. We instead ascribe the lack of quantum Hall states below $\nu = -1$ to the flatness of the underlying bands, such that quantum Hall gaps are suppressed. From our continuum model theory (Supplementary Sec. 3), the bandwidth of the lowest energy moiré band is quite small, with a correspondingly small tight binding model hopping parameter $t = 0.2$ meV. Expanding the dispersion about Gamma gives an effective mass $m^* = \frac{\hbar^2}{2ta^2}$, where $a$ is the lattice constant of the moiré lattice, 5.6 nm for our 3.4$^\circ$ twisted sample. This yields an effective mass $m^*  = 6.1m_e$ where $m_e$ is the electron mass, and a corresponding Landau level gap size of 0.2 meV at $B = 11$ T. This is essentially at our experimental noise floor. While the additional interface in the twisted region could trap disorder beyond that present in the pristine bilayers, which could suppress Landau levels in the twisted region independent of the effective mass of the flat moir\'e bands, the widths of the incompressible peaks in each respective region of the sample are of the same order of magnitude. This suggests that the extrinsic effect of interfacial disorder on Landau levels is minimal.

\begin{figure}[h!]
    \renewcommand{\thefigure}{S\arabic{figure}}
    \centering
    \includegraphics[scale =1]{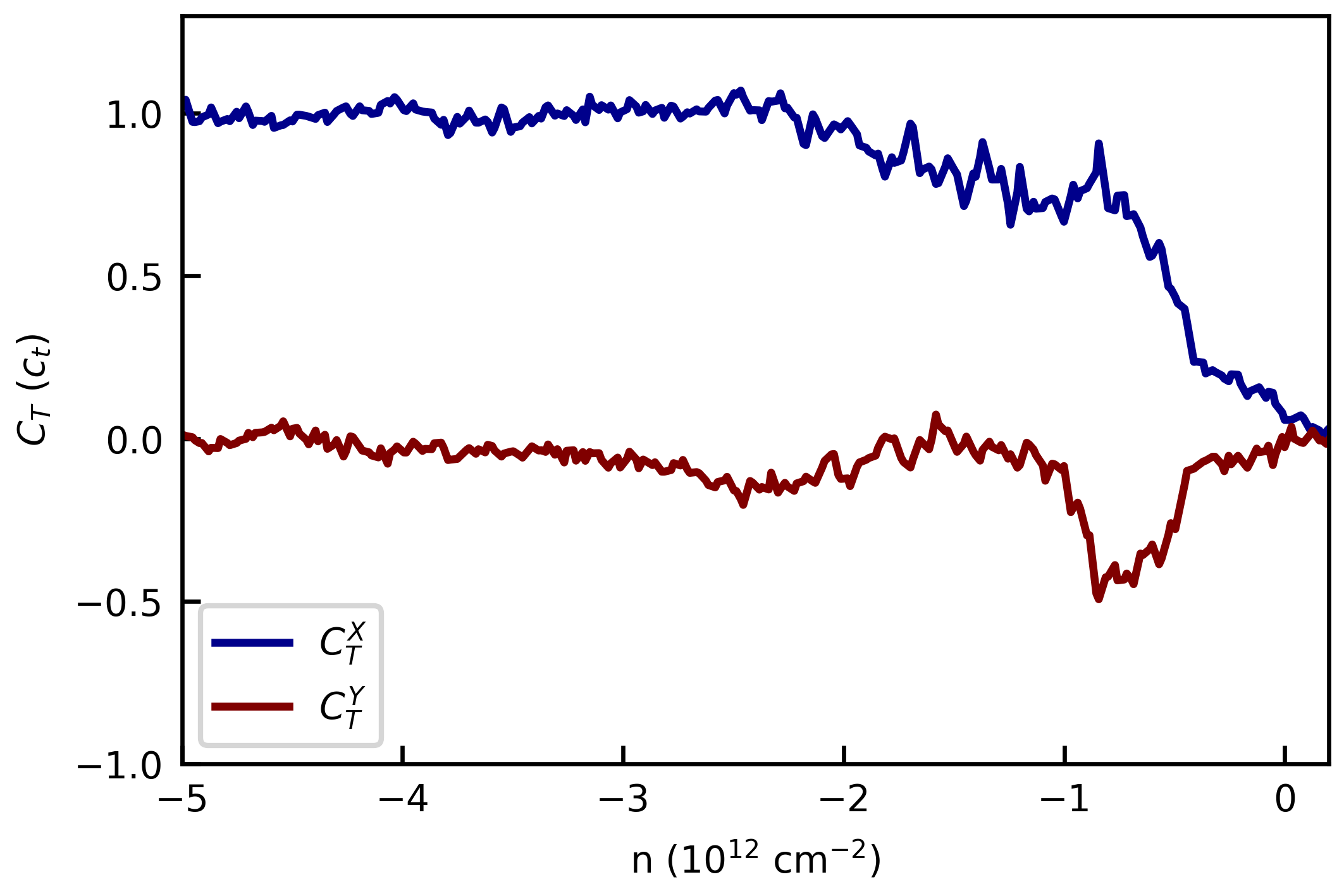}
    \caption{\textbf{Capacitance signal measured at higher contact gate voltages}. A 9 mV a.c. signal is applied to the top gate at a frequency of 911 Hz, and top and back gate voltages are swept to maintain $D = 0$ V/nm.} 
    \label{fig:bandedge}
\end{figure}

In Samples C1/C2, we did not conduct measurements at nonzero magnetic field and thus must rely on the capacitance measurements themselves to estimate the band edge. In Fig. \ref{fig:bandedge}, we show a linetrace of raw capacitance signal taken at a much higher contact gate voltage. This has the effect of improving the sample contact resistance without affecting the device region measured via capacitance to the top gate, as the top gate screens out any stray field from the contact gates which are above it. In most of the measured parameter space, the signal was indeed unaffected. However, it did lead to a nonzero signal extending to lower and lower densities. We take the point where this signal becomes zero within the signal to noise as zero density, and note that this also agrees with the spacing between the $\nu = -1$ and $\nu=-3$ features in Fig. \ref{fig:Fig4} (i.e., the gate voltage range between $\nu = -1$ and $\nu = -3$ is twice that between $\nu = -1$ and the band edge, $\nu = 0$).

\begin{figure}[h!]
    \renewcommand{\thefigure}{S\arabic{figure}}
    \centering
    \includegraphics[scale =1]{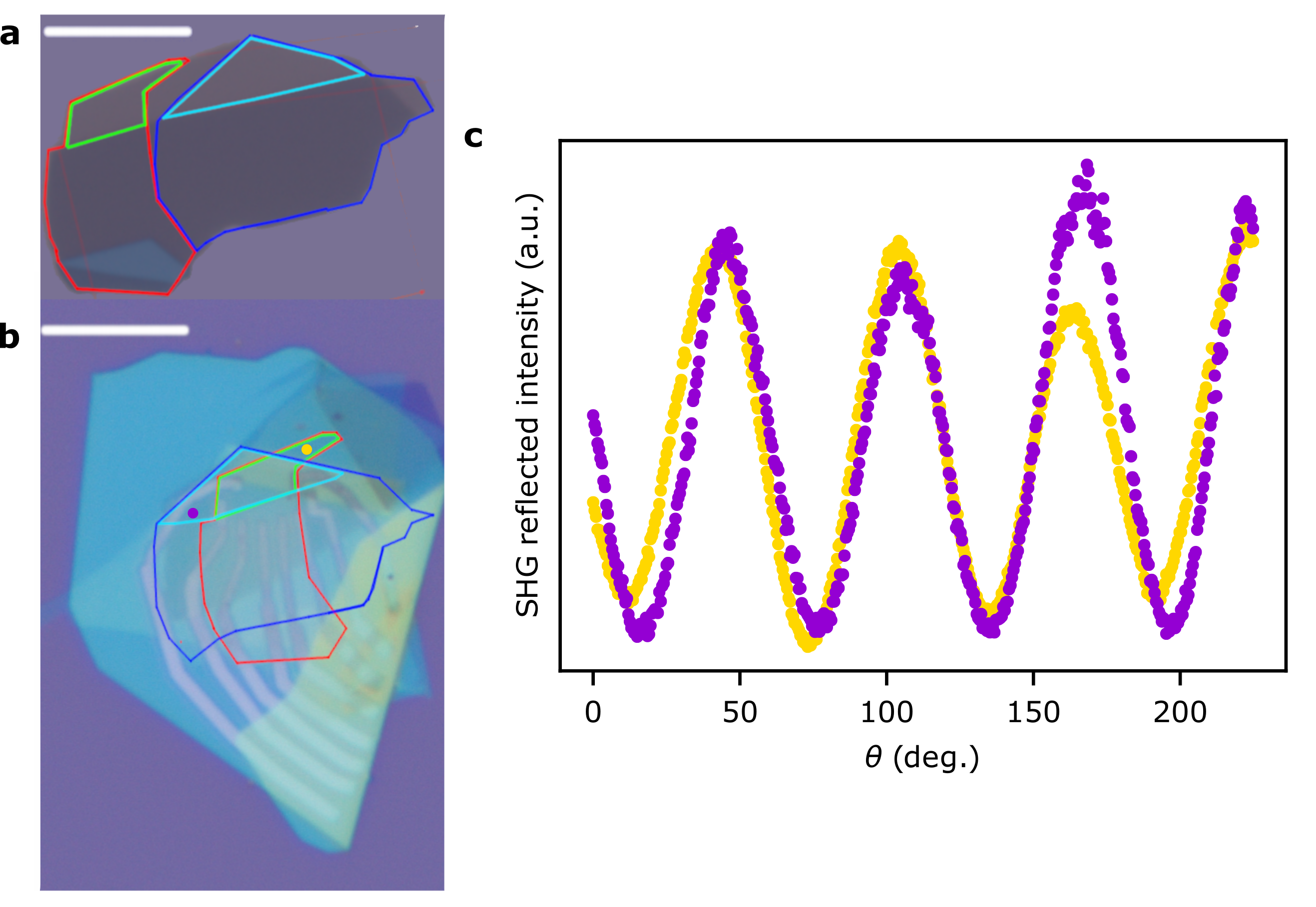}
    \caption{\textbf{Second harmonic generation (SHG) reflectance on Sample S1}.\textbf{a}, Optical image of the WSe$_2$ flake used to stack Sample S1. The two halves respectively are outlined in dark blue and red. The monolayer pieces of each half are outlined in cyan and lime green, with the rest of the darker portion of the flake being bilayer. At the bottom of the flake, a small area of thicker WSe$_2$ is present. The scale bar is 10 $\mu$m. \textbf{b}, Optical image of Sample S1 with the locations of the SHG signal measured in panel \textbf{c} indicated by purple and yellow dots, with the same pieces of the WSe2 flakes outlined as in panel \textbf{a}. The scale bar is 10 $\mu$m.  \textbf{c}, SHG signal as a function of polarization angle $\theta$ on two monolayer regions of Sample S1.}
    \label{fig:SHG}
\end{figure}

We also provide more details confirming the twist angle of all samples studied. A common technique used to measure the twist angle of TMD devices is second harmonic generation (SHG) \cite{wang_correlated_2020,tang_simulation_2020}. However, due to the inversion symmetry of bilayer TMDs, they do not have a SHG response, so that technique for measuring the twist angle is not generally possible in double bilayer samples. Serendipitously, however, the original bilayer WSe$_2$ flake stacked to obtain Sample S1 also had an adjacent monolayer region (Fig. \ref{fig:SHG}a). Portions of this contiguous monolayer area were also twisted and stacked simultaneously with the bilayer parts of the flake, yielding identical twist angle as in the twisted double bilayer regions. We conducted polarization dependent SHG measurements on a monolayer region from each half of the twisted device (yellow and purple dots, Fig. \ref{fig:SHG}b), and obtained a twist angle of 3.61 ± 0.27 degrees by fitting to the oscillations of each signal (Fig. \ref{fig:SHG}c). This represents an independent measure of twist angle, and we obtained similar values throughout the monolayer areas. The quoted value in the text ($3.4^\circ$), which takes the observed incompressible state at $n = -3.9 \times 10^{12}$ cm$^{-2}$ to be $\nu = -1$ indeed falls within this range. 

Other measured samples do not have monolayer regions, so SHG is not informative. However, optical and atomic force microscopy (AFM) images support the twist angle assignments we infer from electronic compressibility measurements for all devices. In Table \ref{tab:twistangles}, we provide additional images of each sample before and after twisting/stacking, and we enumerate three angles. The first angle is the target twist angle during the stacking process, i.e. the angle to which the stage was rotated between picking up the two halves of the twisted device. We also list the angle estimated from optical and AFM images. The images in Table \ref{tab:twistangles} are annotated with blue and red lines that highlight edges of the parent flake which are also clear in the final stacking images. We present optical images because they provide better contrast for the entire TMD flakes, and therefore give more accurate measurements of twist angle. However, analysis of the edges measured with AFM yield consistent results. We estimate the angle from optical/AFM images based on the twist angle required to make both of these edges align with the original flake, with uncertainty estimated based on how sharp the edges are by eye. Finally, we quote the angle determined from the electronic measurements. These are extracted based on the densities at which we observe incompressible states, which we assume to occur at integer moiré filling.

For each sample, the angles all mutually agree within experimental uncertainty. In all cases, the target angle during stacking matches the angle determined from the density at which we observed incompressible states to within $0.3^\circ$. This is comparable to values typically quoted in the literature. In addition, both angles match the range extracted from optical/AFM images within experimental uncertainty. In contrast, alternate filling factor assignments for the observed incompressible states are not consistent with experimental observations. For example, the most compelling and experimentally plausible alternate filling factor assignment for the most prominent incompressible state would be $\nu=-2$, full filling of the lowest energy moiré band. However, the twist angle that would be required for this falls outside the experimental range of twist angle uncertainty for all four devices. For example, the twist angle which would be required for the gap at $n = -3.9\times 10^{12}$ cm$^{-2}$ in Sample S1 to correspond to $\nu = -2$ would be roughly $2.5^\circ$. That angle is well below the range of either the experimentally measured SHG or the estimates of twist angle from optical/AFM images. Finally, we note that the high degree of twist angle homogeneity observed over micron length scales in local scanning SET measurements makes local twist angle variations distinct from the device edges in optical/AFM images highly unlikely.
\begin{table}[]

\renewcommand{\thetable}{S\arabic{table}}

    \centering
    \begin{tabular}{|p{1.3cm}|p{5cm}|p{5cm}|p{2.3cm}|p{2cm}|p{2.3cm}|}
    \hline
        Sample & Image of original flake & Image after twisting and stacking & Target angle (stage rotation during stacking) & Angle (from optical / AFM image) & Angle (from compressibility) \\
    \hline
        S1 & \includegraphics[width=4.5cm,valign=T]{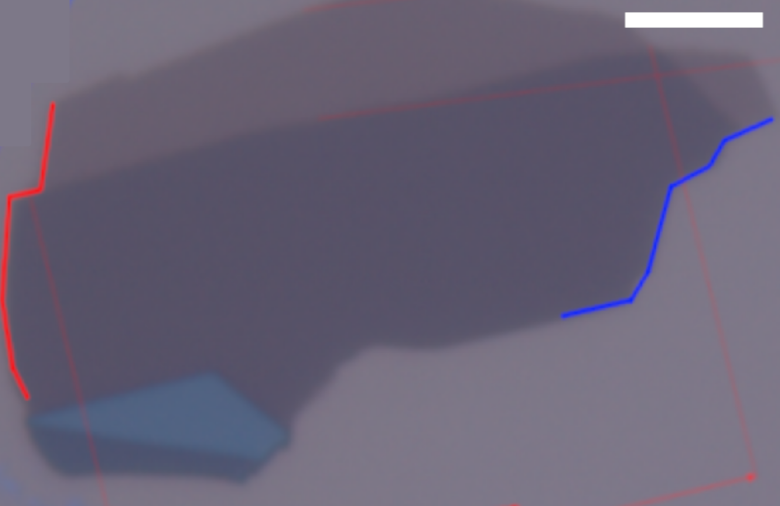} & \includegraphics[width=4.5cm,valign=T]{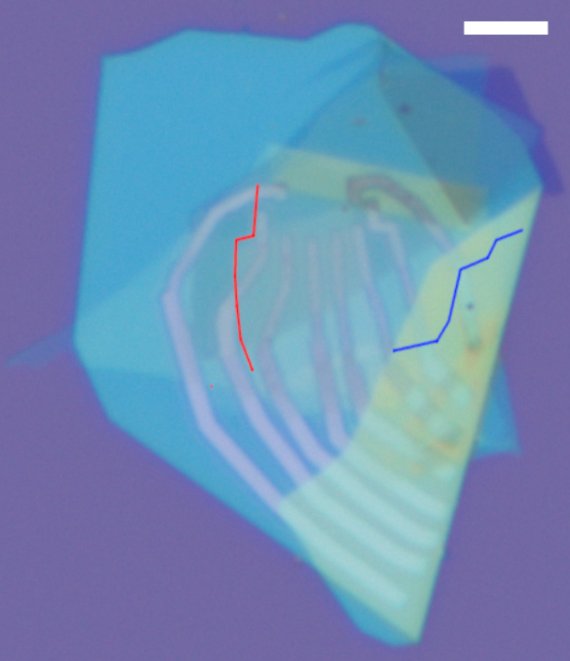} & $3.5^\circ$ & $3.5 \pm 0.3^\circ$ (SHG: $3.6 \pm 0.3^\circ$)& $3.4^\circ$ \\
    \hline
        S2 & \includegraphics[width=4.5cm,valign=T]{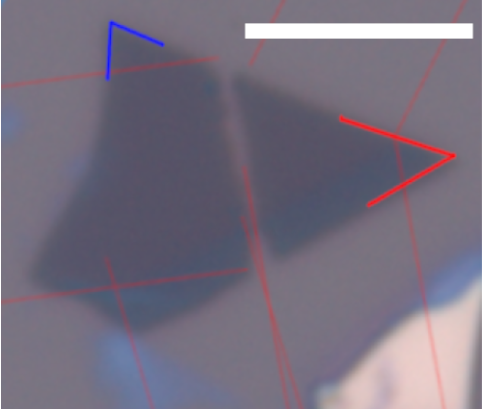} & \includegraphics[width=4.5cm,valign=T]{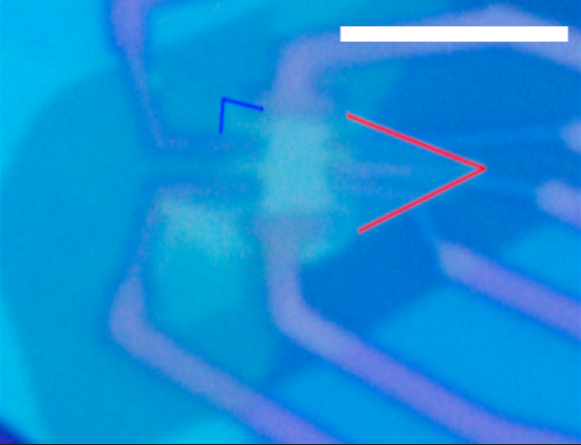} & $2.5^\circ$ & $2.6 \pm 0.4^\circ$ & $2.6^\circ$\\
    \hline
        C1 & \includegraphics[width=4.5cm,valign=T]{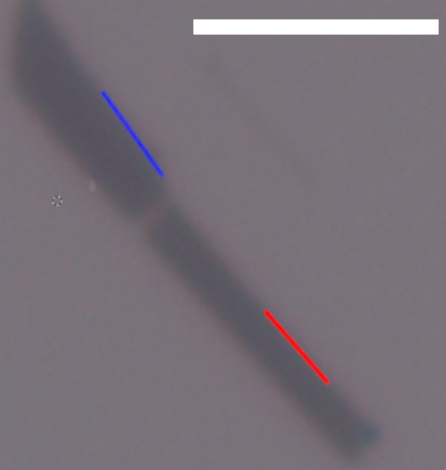} & \includegraphics[width=4.5cm,valign=T]{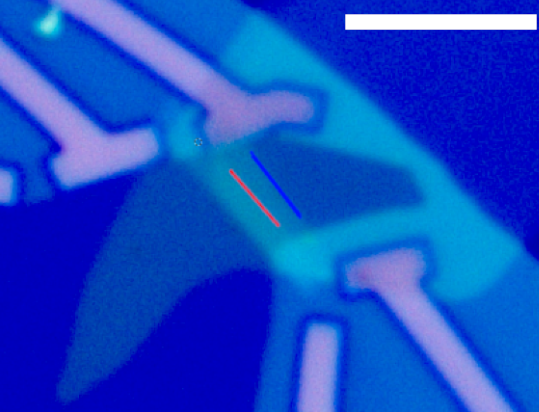} & $2.5^\circ$ & $2.5 \pm 0.7^\circ$ & $2.2^\circ$\\
    \hline
        C2 & \includegraphics[width=4.5cm,valign=T]{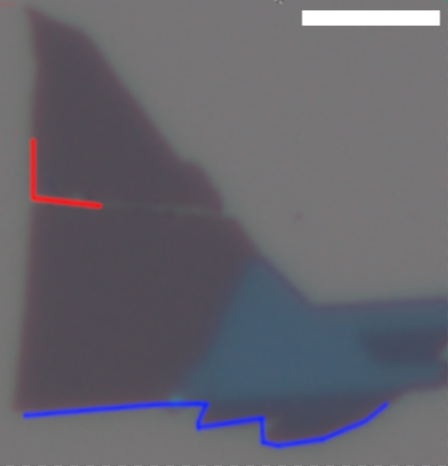} & \includegraphics[width=4.5cm,valign=T]{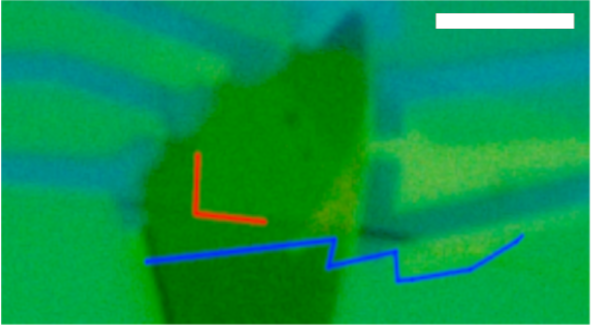} & $3.0^\circ$ & $3.0 \pm 0.5^\circ$ & $3.1^\circ$\\
    \hline

    \end{tabular}
    \caption{ For all samples studied in the manuscript, we provide images of the “parent” bilayer WSe$_2$ flakes and images of the devices once fully stacked/twisted. Using red and blue lines, we highlight clear edges that appear from each half of the flake in the final stacking; these edges are used to estimate the angle from the optical image. Thin red lines on the image of original flakes (on S1 and S2) are due to annotations and should be ignored. For each device, we provide the angle to which we twisted the stage while stacking, the angle estimated from the optical/AFM images, and the angle extracted from compressibility experiments based on the density of observed features. All scale bars are 5 $\mu$m.}
    \label{tab:twistangles}
\end{table}

\section{12. Temperature dependence of capacitance data}
\begin{figure}[h!]
    \renewcommand{\thefigure}{S\arabic{figure}}
    \centering
    \includegraphics[scale =1]{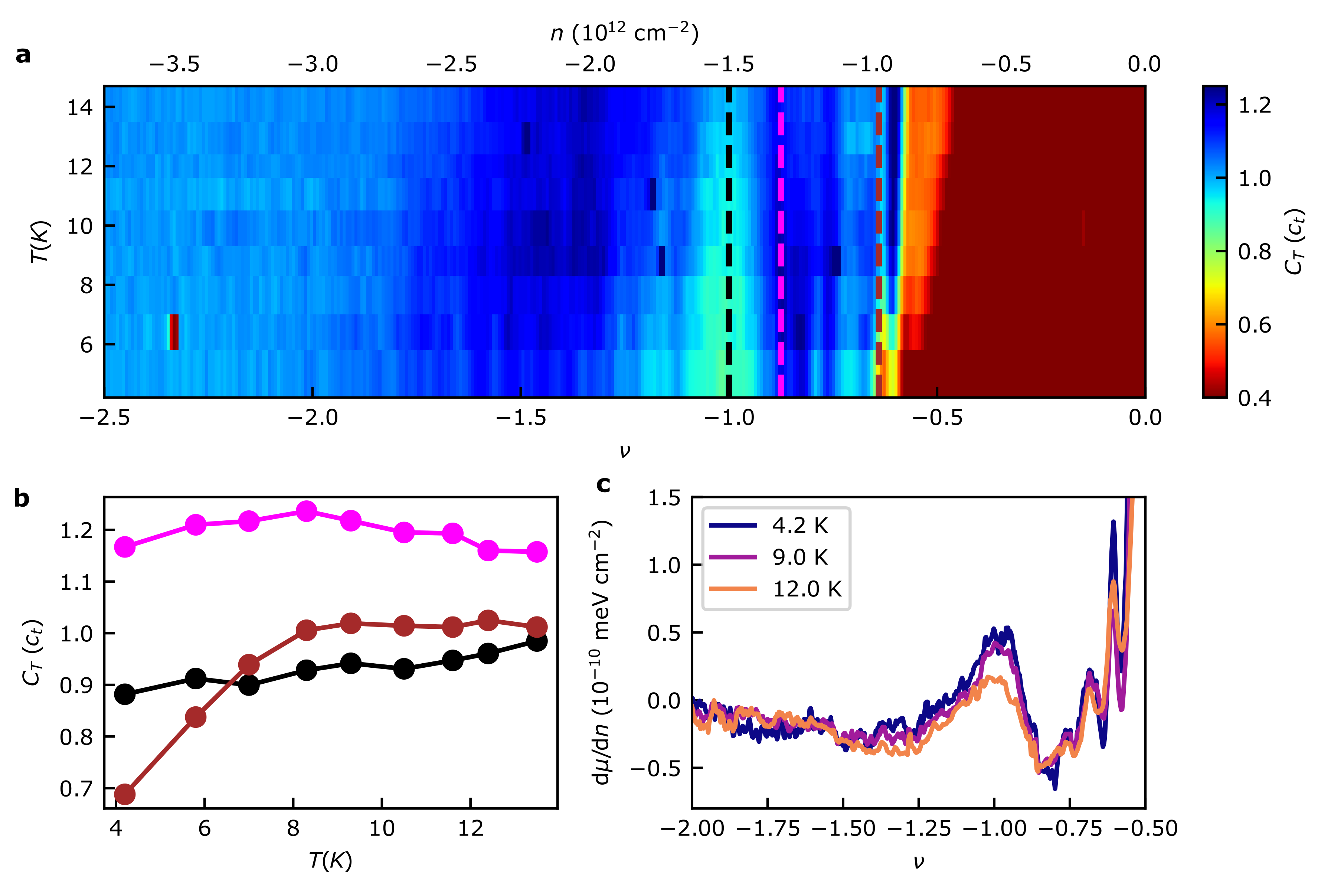}
    \caption{\textbf{Temperature dependence in Sample C1}. \textbf{a}, Temperature dependence of measured capacitance $C_T$ in units of geometric capacitance $c_t$. An 8 mV excitaiton at a frequency of 739.1 Hz and a constant d.c. voltage $-1.6$ V is applied to the top gate. \textbf{b}, Line cuts as a function of temperature from panel \textbf{a} at $\nu = -1$, $\nu = -0.875$, $\nu = -\frac{2}{3}$. \textbf{c} Measured d$\mu$/d$n$ at $D = -0.2$ V/nm at various temperatures.}
    \label{fig:TempLinecuts}
\end{figure}

\begin{figure}[h!]
    \renewcommand{\thefigure}{S\arabic{figure}}
    \centering
    \includegraphics[scale =1]{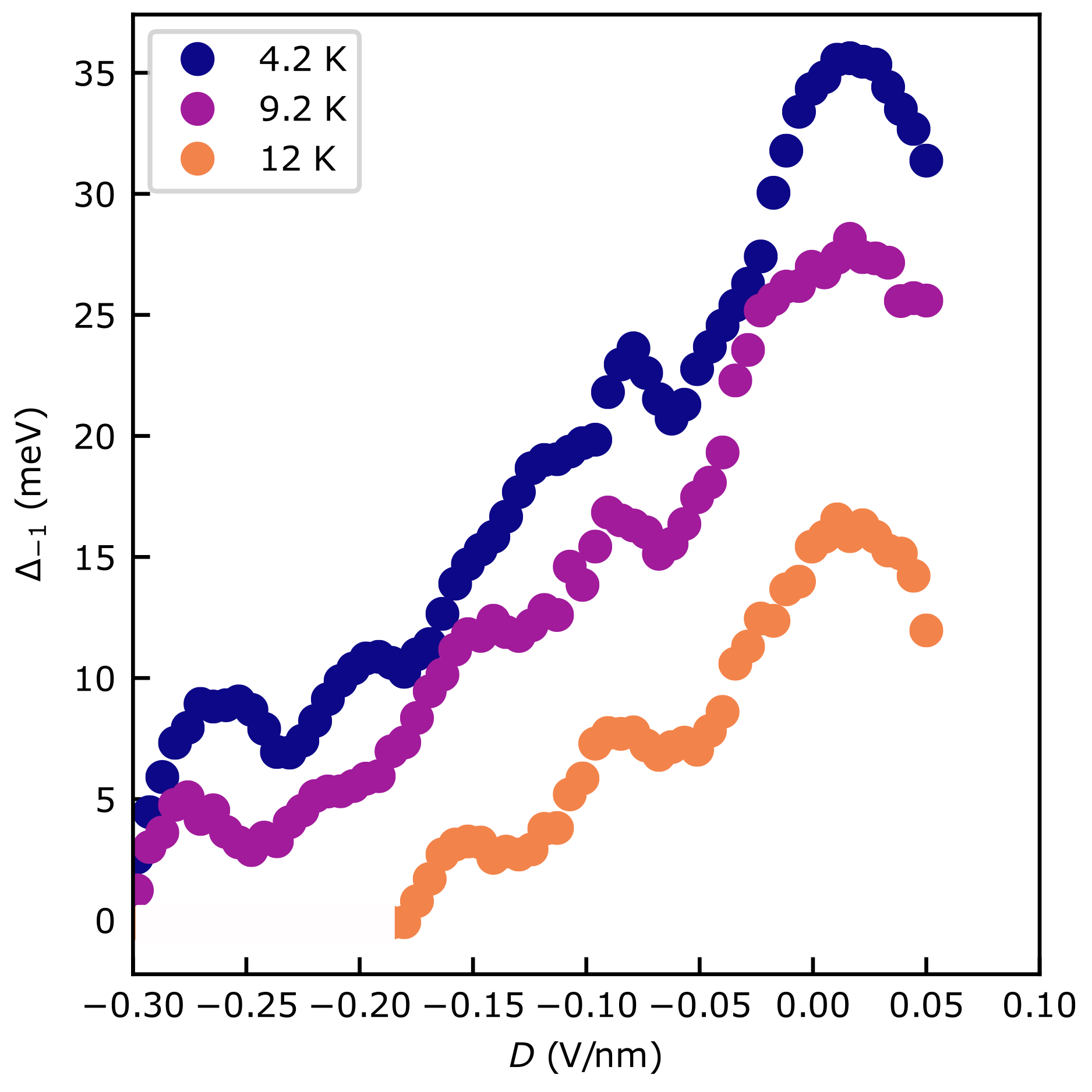}
    \caption{\textbf{Temperature dependence of the $\nu = -1$ gap}. }
    \label{fig:TempGap}
\end{figure}

The temperature dependence of Sample C1 is shown up to $T = 15$ K in Fig. \ref{fig:TempLinecuts}. The magnitude of the incompressible peak at $\nu=-1$ weakens with increasing temperature. In Fig. \ref{fig:TempLinecuts}a-b, $X$-channel (capacitive) data is presented in units of the geometrical top-gate capacitance $c_t$. As the temperature is increased, the charging of the tdWSe$_2$ sample at low densities improves. This is particularly notable at the maroon dashed line ($\nu = -\frac{2}{3}$) where at $T=4.2$ K the capacitance is at $\approx 0.7 c_t$ but it quickly rises as the mobility edge extends rightwards. A more modest version of this effect is also seen at $\nu = -1$ (black), whereas in the negative compressibility region (pink), the temperature dependence is nonmonotonic, first increasing (presumably) due to better charging before decreasing at the highest temperatures. In Fig. \ref{fig:TempLinecuts}c, we show line cuts at constant $D = -0.2$ V/nm at a few temperatures. The incompressible peak at $\nu = -1$ is weaker at 12 K relative to 4.2 K, but still clearly evident. In Fig. \ref{fig:TempGap} we show the behavior of the $\nu = -1$ gap as a function of $D$ at a few distinct temperatures. As in previous reports, the gaps show a linear decrease in gap size as the temperature increases, likely due to thermal broadening \cite{li_charge-order-enhanced_2021}. The slope of gap size $\Delta_{-1}$ with $D$ is similar across various temperatures.

\section{13. Estimation of quantum capacitance from global capacitance}
\begin{figure}[h]
    \renewcommand{\thefigure}{S\arabic{figure}}
    \centering
    \includegraphics[scale =1]{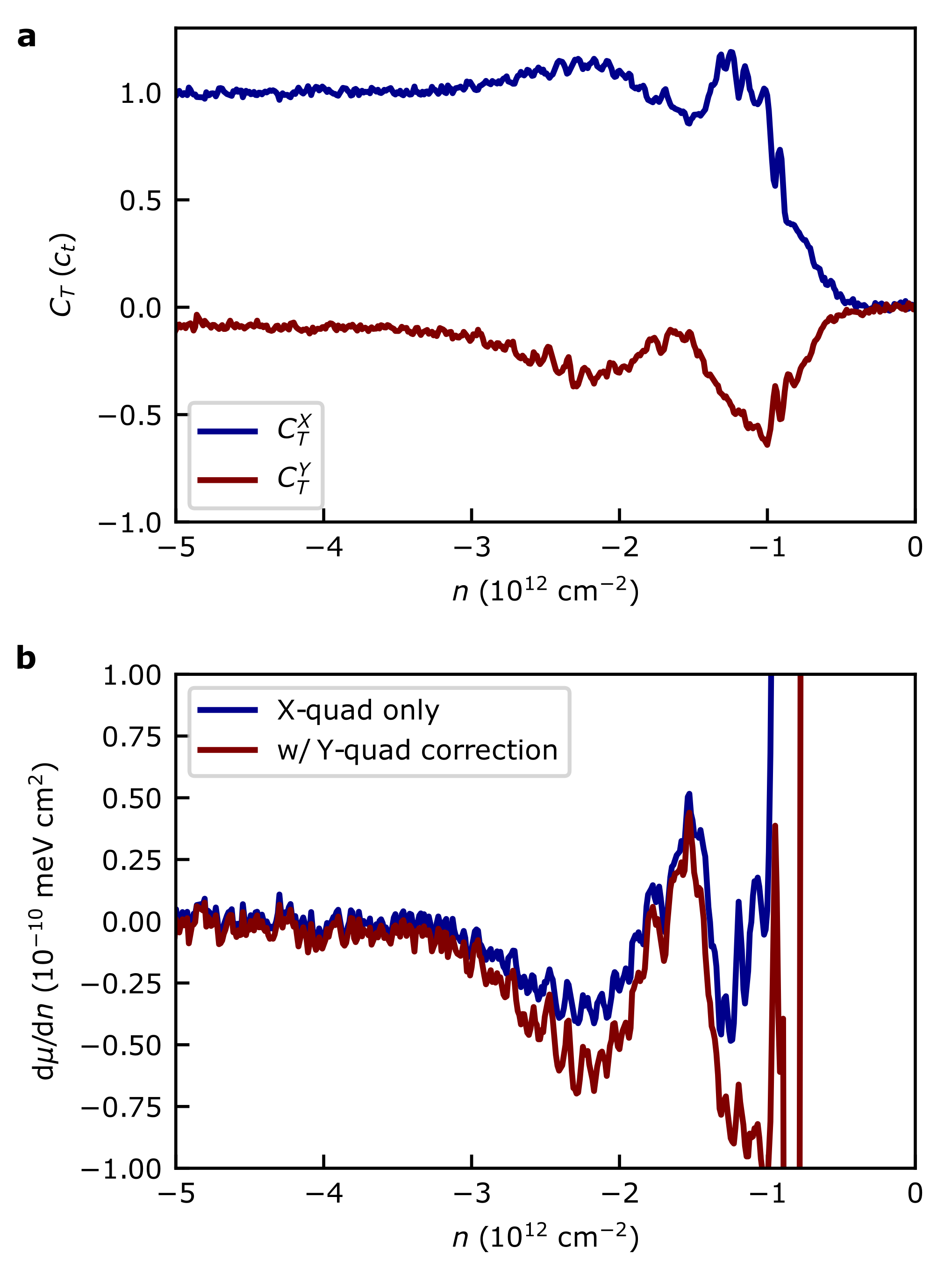}
    \caption{\textbf{Example capacitance measurements on Sample C1}. \textbf{a}, Line cuts showing the capacitance signal in the $X$ (capacitance) and $Y$ (dissipation) channels. \textbf{b}, Extracted $d\mu/dn$ signal using each respective method discussed in the text. Data were measured at a temperature of 4.2 K with an 8 mV excitaiton applied to the top gate at frequency 739.1 Hz and a constant d.c. voltage $-1.6$ V on the top gate.} 
    \label{fig:dudnmodel}
\end{figure}

As discussed in the Methods, we convert the capacitance signal into units of the geometric capacitance $c_t$ by assuming $C_T = 0$ in the gap between the WSe$_2$ valence and conduction bands and $C_T = c_t$ at high hole dopings $>5\times 10^{12} \textrm{cm}^{-2}$. Using a lumped circuit capacitance model \cite{li_charge-order-enhanced_2021,shi_odd-_2020}, the measured top gate capacitance $C_T$ is given by 
\begin{equation}\label{eq:1}C_T = \frac{c_tc_q}{c_t+c_b+c_q} \end{equation} where $c_t$ and $c_b$ are the geometric top and bottom gate capacitances and $c_q \equiv e^2 (d\mu/dn)^{-1}$ is the quantum capacitance. As discussed in the main text, this approximation will overestimate $d\mu/dn$ in the case of high sample resistance or high frequency. Some of this enhancement can be accounted for by including the simultaneous measurement of the out-of-phase quadrature of the measured capacitance, which is equivalent to the dissipation signal \cite{goodall_capacitance_1985,shi_bilayer_2022}. In this case, the total measured signal is given by \begin{equation}
    V^{\textrm{complex}}\propto C_T/c_t = \frac{c_q}{c_t+c_b+c_q} \frac{\tanh(\alpha)}{\alpha}
    \label{eq:complex1}
\end{equation}
\begin{equation}
    \alpha = \sqrt{i\omega \frac{c_qc_t}{c_t+c_b+c_q} R_s}
    \label{eq:complex2}
\end{equation}
where $\omega$ is the measurement frequency and $R_s$ is the sample resistance. Note that in the small $\alpha$ limit, the imaginary (capacitive) part of this signal agrees with Eq. \ref{eq:1} and the real (dissipative) part is zero. This system can be solved for $c_q$ and $R_s$ using both the $x$ and $y$ quadratures, and the resulting $c_q$ yields $d\mu/dn$ as before. An example of $x$ and $y$ quadrature capacitance signal and the extracted $d\mu/dn$ by both of these methods are shown in Fig. \ref{fig:dudnmodel}. All data presented from capacitance measurements in the main text show $d\mu/dn$ calculated using this correction. The main effect of including the dissipation on the extracted $d\mu/dn$ occurs in the regions of parameter space  which exhibit negative compressibility; there is also a consistent small decrease of the broad peak at $\nu = -1$ (Fig. \ref{fig:dudnmodel}). 

\begin{figure}[h]
    \renewcommand{\thefigure}{S\arabic{figure}}
    \centering
    \includegraphics[scale =1]{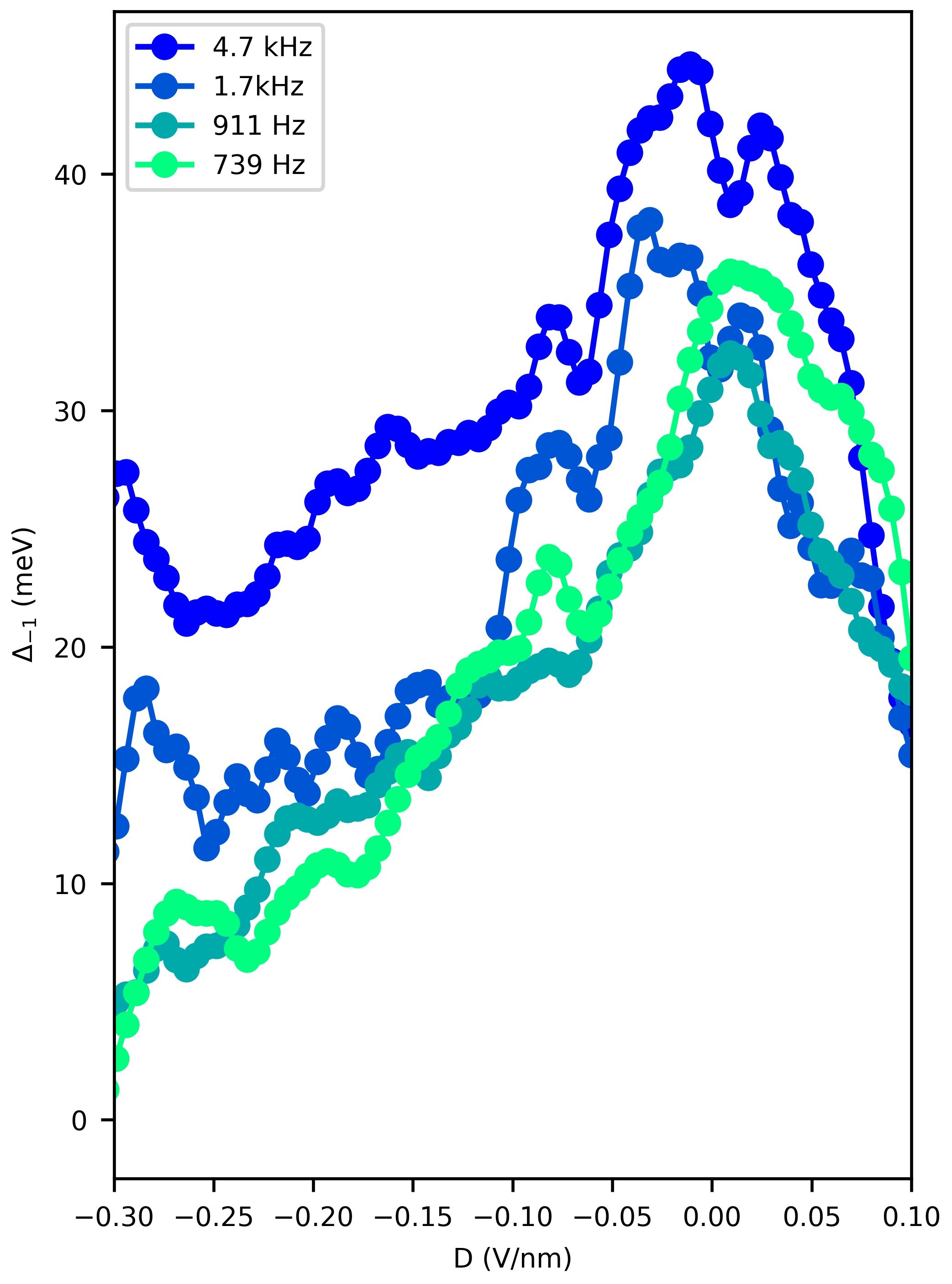}
    \caption{\textbf{Frequency dependence}. Measured gap $\Delta_{-1}$ in Sample C1 at various frequencies.} 
    \label{fig:gapfreq}
\end{figure}

In Fig. \ref{fig:gapfreq} we show a comparison of the measured charge gap at $\nu = -1$ as a function of applied displacement field $D$ at various frequencies. At 4.7 kHz, the measured gap is clearly enhanced above the lower frequency curves. Below 1.7 kHz and 739 Hz, there is little measured change. This lack of change suggests that the extraction of $d\mu/dn$ from Eqs. \ref{eq:complex1} and \ref{eq:complex2} performs adequately at these measurement frequencies.

%